\documentclass[letterpaper,journal,final]{IEEEtran}
\IEEEoverridecommandlockouts

\newcommand{\subparagraph}{}

\usepackage{graphicx,psfrag,dblfloatfix}
\usepackage{amsmath,amssymb,amsfonts,mathrsfs}
\usepackage{color,epstopdf}
\usepackage{algorithmic}
\usepackage{enumerate} 
\usepackage{subfigure}
\usepackage{float}
\usepackage{flushend}
\newtheorem{remark}{Remark}
\newtheorem{theorem}{Theorem}
\newtheorem{lemma}{Lemma}

\newcommand{\eps}{\varepsilon}
\newcommand{\neigh}[1]{{\cal N}_{#1}}
\newcommand{\ave}{\operatorname{ave}} % 
\newcommand{\sign}{\operatorname{sign}} %

\def\qedp{\hspace*{\fill}~{\tiny $\blacksquare$}}
\def\be{\begin{equation}}
\def\ee{\end{equation}}
\def\ba{\begin{array}}
\def\ea{\end{array}}
\def\eqa{\begin{eqnarray}}
\def\eqe{\end{eqnarray}}

\definecolor{darkgreen}{rgb}{0.0, 0.55, 0.0}
\definecolor{amaranth}{rgb}{0.9, 0.17, 0.31}
\usepackage[prependcaption,colorinlistoftodos]{todonotes}

\begin{document}

\title{
%\textcolor{red}{
Self-Triggered Network Coordination over Noisy Communication Channels}
%}

\author{M. Shi, P. Tesi   and {C. De Persis}
\thanks{M. Shi, P. Tesi   and {C. De Persis}  are with ENTEG, 
University of Groningen, 9747 AG Groningen, The Netherlands.
Email: {\tt\small M.Shi@rug.nl, p.tesi@rug.nl, c.de.persis@rug.nl.}
P. Tesi is also with DINFO, University of Florence, 50139 Firenze, Italy 
E-mail: {\tt\small pietro.tesi@unifi.it}.}
}

\maketitle
\begin{abstract}
This paper investigates coordination problems over packet-based communication channels.
We consider the scenario in which
the communication between network nodes is corrupted by unknown-but-bounded noise. 
We introduce a novel coordination scheme, which
ensures practical consensus in the noiseless case, while preserving 
bounds on the nodes disagreement in the noisy case.
The proposed scheme does not require any global information 
about the network parameters and/or the operating environment (the noise characteristics). 
Moreover, network nodes can sample at independent rates and in an aperiodic manner.
The analysis is substantiated by extensive numerical simulations.
\end{abstract}

%===============================================================================

\section{%\textcolor{blue}{
Introduction
%}
}

\IEEEPARstart{D}{istributed} 
coordination is one of the most active research areas in control engineering,
with applications ranging from sensor fusion to 
optimization and control \cite{cao2013overview}. 
To achieve coordination, the network units (nodes) 
have to collect and process data from   
neighbouring nodes. In practice, a main issue is that
the data transmission is often carried out through digital (\emph{i.e.}, packet-based) 
communication channels. 
Thus, coordination 
algorithms should take into account that the data 
exchange can only occur at finite rates
and that the communication medium can introduce issues 
such as packet loss, transmission delay and {noise}. 

The goal of this paper is to study coordination algorithms 
in the presence of communication noise, which is one of the 
major issues that arise in problems involving data exchange. 
We shall focus on \emph{consensus} \cite{cao2013overview} algorithms since 
consensus is the prototypical problem in distributed coordination. \smallskip

\emph{Literature review}

Even if one neglects network-related issues such as finite transmission rate, dropouts and 
delay, developing noise-robust consensus algorithms
is a very challenging task. The intuitive reason is that consensus algorithms usually 
rely on Laplacian dynamics.
Since the Laplacian matrix is only marginally stable 
(has an eigenvalue at zero), communication noise  
can cause the state {of} the nodes to diverge. This means that
even if consensus is achieved the consensus value need not be bounded,
in which case convergence may become useless.

Most of the research works in this area assume that the noise
has specific statistical properties, for example that it is 
\emph{white} \cite{LiAUT2009,cheng2011necessary}, \emph{Brownian}-like \cite{li2014multi}
or \emph{martingale} \cite{li2010consensus,huang2010stochastic}.
In contrast, only few research works have approached the problem where,
due to uncertain channel characteristics,
one can only regard noise as a bounded signal (\emph{unknown-but-bounded}).
Arguably, the lack of noise statistical properties makes it much more
difficult to ensure state boundedness since one cannot rely on features 
such as as \emph{zero-mean} or \emph{stationarity}.
In \cite{kingston2005consensus}, 
the authors consider a Kalman-based coordination scheme, 
and show that consensus (the disagreement among the nodes) satisfies input-to-state stability properties,
but no results are given regarding boundedness of the state trajectories.
In \cite{shi2013robust}, the authors study 
robust and integral robust consensus with respect to 
$L_\infty$ and $L_1$ norms of the noise function, 
but again the analysis only involves the disagreement variable  
and no results are given regarding state boundedness. This is also the 
case in \cite{garulli2011analysis} where the authors consider 
discrete consensus under bounded measurement noise, and in \cite{FranceschelliNAHS2013}
where the authors propose discontinuous interaction rules to mitigate 
the effect of disturbances on the nodes disagreement.
In the slightly different context of leader-following consensus, 
\cite{arabi2017mitigating} considers the issue of sensor noise, 
but assumes that the noise is a smooth signal. While effective to account 
for sensor bias, this hypothesis is hardly met with 
communication noise. A framework {closer} to ours is in 
\cite{bauso2009consensus}. There,
the authors propose a coordination scheme that guarantees \emph{approximate} consensus 
along with boundedness of the state trajectories,
but an upper bound on the magnitude of the noise is required to be known.
\smallskip

\emph{Summary of contributions}

In this paper, we consider a novel coordination algorithm that can handle
\emph{unknown-but-bounded} noise without requiring the knowledge of a noise upper bound.
In order to prevent state divergence, we propose a \emph{state-dependent} coordination scheme
where each node dynamically adjusts its update rule depending on the magnitude of its state.
This approach can be regarded as a coarse dynamic quantization strategy, 
which updates the quantization based on the state of the nodes \cite{carli2010quantized}. 
We show that this approach prevents state divergence and guarantees,
in the noiseless case, a maximum consensus error for the worst case over the initial vector of states,
which is reminiscent of \emph{normalized} consensus metrics \cite{boyd2006randomized,dimakis2010gossip}. 
As for the noisy case, we show that this approach guarantees that both disagreement and 
state variables scale nicely (linearly) with the noise magnitude.

From a technical point of view, our approach employs a \emph{self-triggered} control scheme \cite{de2013robust}. 
Each node uses a local clock to decide its update times. 
At each update time, the node polls its neighbors, collects the data and determines 
whether it is necessary to modify its controls along with its next update time.
Similar to \emph{event-triggered} control 
\cite{heemels2012introduction}\nocite{imarogonas2012distributed}-\cite{nowzari2017event},
\emph{self-triggered} control 
\cite{anta2010sample}\nocite{tolic2013,fan2015self}-{\cite{depersisTAC2017}}
features the remarkable property
that the communication among nodes occurs only
at discrete time instants. Moreover, the nodes can sample independently and aperiodically.
Thus, the proposed approach is appealing also from the perspective of finding coordination algorithms  
that are practically implementable
(as we will see, including the case where the data exchange encounters delays).  

The proposed self-triggered algorithm shares similarities with 
several pairwise gossip or multi-gossip approaches with  randomized \cite{Boydrandom} 
and deterministic \cite{JiliuDeterminisitc} protocols. 
There is however a major difference, namely that while for gossiping algorithms 
the inter-node interaction times occur at multiples of discrete time-steps, 
in self-triggered consensus algorithms the update instants are established 
on the basis of current node measurements and can take any value on the continuous-time axis. 
Moreover, to the best of our knowledge, gossiping has not been considered 
in connection with unknown-but-bounded noise, even in the recent literature 
\cite{GDShiFiniteT}-\cite{ChangbinyuPeriod2017}. 

A preliminary version of the manuscript appeared in \cite{STPnoisy}. Compared with the latter, this paper provides complete proofs of all the results, a thorough discussion of the  proposed method and extensive numerical results. Furthermore a new section considering the presence of delays in the communication channel is considered.

The remainder of the paper is as follows. In Section \ref{sec:PM}, 
we introduce the framework of interest 
and outline the main paper results. The proposed 
coordination scheme is introduced in Section \ref{sec:scheme},
and the main results are discussed throughout 
Sections \ref{sec:noiselesscase} and \ref{sec:Noisycase}. 
Section \ref{sec:n2n_error} further discusses properties of 
the proposed coordination scheme. Numerical examples are illustrated in
Section \ref{sec:examp}. Section \ref{sec:conclusion} ends the paper with concluding remarks. 
In the Appendix, the main results of the paper are extended 
to include communication delays.

\subsection{Notation}

We assume to have a set of nodes $I=\{1,2,...,n\}$ and an
undirected connected graph $G=(I,E)$,
where $E \subseteq I \times I$ is the set of edges (links).
We denote by $L$ the Laplacian matrix of $G$, which is a symmetric matrix. 
For each node $i\in I$, we denote by ${\cal N}_{i}$ the set of 
its neighbors and by $d_i$ its degree, that is, the cardinality of 
${\cal N}_{i}$. Given $n$ scalar-valued variables $v_1,v_2,\ldots,v_n$,
we define $v:=\textrm{col}(v_1,v_2,\ldots,v_n)$.
Given a vector $v \in \mathbb R^n$, $|v|$ denotes its Euclidean norm and $|v|_\infty$ its infinity norm.
Given a signal $s$ mapping $\mathbb R_{\geq 0}$  to $\mathbb R^n$, we define
$|s|_\infty := \sup_{t \in \mathbb R_{\geq 0}} |s(t)|_\infty$
and say that $s$ is \emph{bounded} if $|s|_\infty$ is finite.

\section{Framework and Outline of the Main Results} \label{sec:PM}
 
\subsection{Network dynamics}

We consider a network of $n$ dynamical systems that are 
interconnected over an undirected graph 
$G=(I,E)$. Each node is 
described by
\begin{eqnarray}
\label{eq:modelloA-cont} 
\begin{array}{l}
\dot x_i=u_i\\
z_i=x_i + w_i
\end{array} 
\end{eqnarray}
where $i \in I$; $x_i \in \mathbb R$ is the state;
$u_i \in \mathbb R$ is the control input, and $z_i \in \mathbb R$
is the output where  $w_i \in \mathbb R$ is a bounded signal, which models
communication noise. 
Note that this model implies that all the neighbors of node $i$ will
receive the same corrupted information. As it will become clear in the sequel,
it is possible to replace the second of (\ref{eq:modelloA-cont})
with $z_{ij}=x_i + w_{ij}$, where $i \in I$ and $j \in \mathcal N_i$, so that each neighbor of node $i$
receives a different corrupted information. We will not pursue this model in order to 
keep the notation as streamlined as possible.

According to the usual notion of consensus \cite{cao2013overview}, the network nodes  
should converge, asymptotically or in a finite time, 
to an equilibrium point where all the nodes have the same value lying 
somewhere between the minimum and maximum of their initial values.
In the presence of noise, however, convergence to an exact common value
is in general impossible to achieve. As outlined hereafter, the main contribution of this paper is 
a new coordination scheme that ensures \emph{practical} (\emph{approximate}) consensus, 
namely convergence to a set whose radius depends on the noise amplitude. 

\subsection{Outline of the main results}

One way to define 
practical consensus is via the \textit{normalized} error between the nodes. 
We consider a coordination scheme that, in the noiseless case, 
guarantees that all the network nodes remain
between the minimum and the maximum of their initial values,
and converge in a finite time to a point belonging to the set
\begin{equation}\label{eq:setE}
{\cal E} :=\{x\in \mathbb{R}^n: 
|\sum_{j\in {\cal {N}}_i}(x_j-x_i)| < \max\{\eps, \eps \chi_{0} \},\ \forall i\in I \} 
\end{equation}
where $\eps \in (0,1)$ is a design parameter, and $\chi_0:= |x_i(0)|_\infty$.
In words, when $\chi_0>1$ the coordination scheme guarantees 
that, in a finite time, each node reaches a local average that satisfies 
\begin{eqnarray}
\frac{|\sum_{j\in {\cal {N}}_i}(x_j-x_i)|}{\chi_0} \le\eps
\end{eqnarray}
The parameter $\eps$ determines the desired accuracy level for the consensus 
final value, which is normalized to the magnitude of the initial data.  
In this way, a maximum error $\eps$ is guaranteed for the \emph{worst case} 
over the initial vector of measurements.
If instead $\chi_0\le 1$ then the tolerance reduces to $\eps$. 
We will further comment on this point in Section \ref{sec:n2n_error}.

As for the noisy case, the coordination scheme guarantees that the error scales nicely 
with respect to the noise magnitude. Specifically, let
\begin{eqnarray} \label{eq:r}
r:=\max\{\eps,\eps\chi_0\}+\left(\frac{\eps}{3}+3d_{max}\right)
|w|_\infty
\end{eqnarray}
where $d_{max}:= |d|_\infty$ denotes the maximum among the nodes degrees.
The scheme guarantees that, in a finite time, the network state enters the set
\begin{eqnarray}\label{eq:setD}
{\cal D}:=\{x\in\mathbb{R}^n: |\sum_{j\in {\cal {N}}_i}(x_j-x_i)| < r,\ \forall i\in I\}
\end{eqnarray} 
and remains there forever with convergence in the event that $w$ goes to zero. Moreover, 
the state remains confined in a set whose radius depends on $\eps$ and $|w|_\infty$.

From an implementation point of view, the proposed scheme enjoys the following features:
\begin{enumerate}[(i)]
	\item[(i)] 
	No knowledge of $\chi_0$ is required.
	\item[(ii)] No knowledge of $|w|_\infty$ is required. 
	\item[(iii)] The control action is fully distributed.
	\item[(iv)] The communication between network nodes 
	occurs only at discrete time instants. Moreover, the nodes can sample independently 
	and in an aperiodic manner.
\end{enumerate}
These features indicate the implementation does not require any global information 
about the network parameters and/or the operating environment (the noise). 
The last feature renders the proposed scheme applicable when coordination 
is through packet-based communication networks.

The main derivations will be carried out assuming that there are no communication delays,
which are dealt with in the Appendix section. 
The analysis shows that, in practice, delays have the same effect
as an additional noise source. For this reason, also numerical simulations 
will be restricted to the delay-free case.

\section{Self-triggered Coordination with Adaptive Consensus Thresholds}
\label{sec:scheme}

 \subsection{Adaptive consensus thresholds}

As discussed in the previous section, 
we aim at considering a \emph{normalized} error 
between network nodes. To this end, each node
has a local variable
\begin{eqnarray}\label{eq:vareps}
\eps_i(t) := 
\def\arraystretch{1.2}
\left\{ \begin{array}{ll}
\eps |x_i(t)| & \quad \text{if}\ |x_i(t)| \geq 1 \\ 
\eps & \quad \text{otherwise}
\end{array} \right.
\end{eqnarray}
that specifies the threshold used to assess
whether or not consensus is achieved. 
In contrast with previous  
self-triggered schemes \cite{de2013robust,senejohnny2017jamming},
this threshold is \emph{adaptive} as it scales dynamically with the 
state magnitude. It is exactly this feature that ensures robustness against noise.
 
Notice that $x_i$ 
is used by node $i$ to construct the threshold $\eps_i$, which 
amounts to assuming that each node has access to its own state without noise.
This assumption can be relaxed and all the results continue to hold with the difference that the state 
bound (\ref{thm:statebounded}) and the consensus set radius ($r$ in equation (\ref{eq:r}))
will be enlarged. We neglect the details for this situation since it does not affect the general idea of the paper.

\subsection{Control action and triggering times}

For each $i \in I$, let $\{t_k^i\}_{k \in \mathbb N_0}$ with $t^i_0=0$ be
the sequence of time instants at which node $i$ collects
data  from its neighbors. At these time instants, the node
updates its control action and determines when the next update will be triggered.

For each $i \in I$, let
\begin{eqnarray} \label{eq:ave_noisy}
\ave^w_i(t) := \sum_{j\in\neigh{i}}(z_j(t)-x_i(t))
\end{eqnarray}
denote the local noisy average.

The control action makes use of a quantized $\sign$ function,
The control signals take values in the set $\ \mathcal U : = \{-1,0,+1\}$,
and the specific quantizer of choice is 
$\sign_\alpha : \mathbb R \rightarrow \mathcal U$, $\alpha >0$, which is given by
\begin{eqnarray} \label{eq:signeps}
\sign_\alpha(z) :=
\left\{
\begin{aligned}
&\sign(z)\qquad \textrm{if } |z| \ge \alpha \\ 
& 0 \qquad\qquad\ \; \textrm{otherwise}
\end{aligned} \right.
\end{eqnarray}
The control action is given by
\begin{eqnarray} 
\label{eq:controls}
u_i(t) = \sign_{\eps_i(t^i_k)}\!\left(\ave^w_i(t^i_k)\right) 
\end{eqnarray}
for $t \in [t^i_k,t^i_{k+1}[$.

The triggering times are given by
$t^i_{k+1} = t^i_{k} + \Delta^i_k$, where 
\begin{eqnarray} \label{eq:sam}
\Delta^i_k:=
\def\arraystretch{2.2}
\left\{
\begin{array}{ll}
\displaystyle \frac{|\ave^w_i(t^i_{k})|}{4d_{i}} & \quad \textrm{if } \, 
|\ave^w_i(t^i_{k})|\ge \eps_i(t^i_{k})  \\
\quad\ \displaystyle \frac{\eps}{4d_{i}} & \quad
\textrm{otherwise} 
\end{array} \right.
\end{eqnarray}
Note that by construction the inter-sampling times 
are bounded away from zero. This guarantees the 
existence of a unique Carath\`{e}odory solution
for the state trajectories.

\begin{remark}
In the noise-free case, the control law (\ref{eq:controls}) is an approximation of the pure (non-quantized)
sign function which yields ``max-min" consensus \cite{cortes2006finite}, that is convergence 
to the centre of the interval
containing the nodes initial values. Specifically, in the noise-free case, the scheme
reduces to the one in \cite{cortes2006finite} when $\eps_i(\cdot) \equiv 0$ and the flow of information
among nodes is continuous. We refer the reader to Sections VII-B 
for further discussions on this point. \qedp
\end{remark}
\begin{remark}
Although the paper focuses on networks of dynamical systems 
of the form \eqref{eq:modelloA-cont}, 
it is not hard to tackle  synchronization problems involving linear 
dynamics as in \cite{SCARDOVIsynidentical}, since synchronization
can be reduced to a consensus problem by means of suitable coordinate 
transformations. For the noise-free case self-triggered algorithms 
for the synchronization of linear systems have been studied in \cite{de2013sync}, 
and for the noise-free case with packet dropouts in \cite{DS-PT-CDP-CDC16}. 
These algorithms can be modified in the spirit of (\ref{eq:vareps})-(\ref{eq:sam}) 
for the case of noisy measurements and the analysis carried out in the 
rest of the paper can be extended to the synchronization problem of linear systems. \qedp
\end{remark}

\section
{Noiseless Case}
\label{sec:noiselesscase}

We start by investigating the properties of this coordination scheme in the absence 
of communication noise. 
For ease of notation, we let
\begin{eqnarray} \label{eq:ave_noiseless}
\ave_i(t) := \sum_{j\in\neigh{i}}(x_j(t)-x_i(t))
\end{eqnarray} 
denote the noiseless average. Note that in the noiseless case 
$\ave_i^w(t)=\ave_i(t)$ for every $t \in \mathbb R_{\geq 0}$.

Let 
\begin{eqnarray} 
\overline x := \max_{i \in I} x_i(0), \quad \underline x := \min_{i \in I} x_i(0)
\end{eqnarray} 
We have the following result. \smallskip

\begin{theorem}\label{thm:noiseless}
Consider a network of $n$ dynamical systems as in (\ref{eq:modelloA-cont}) with $w \equiv 0$, which are
interconnected over an undirected connected graph $G = (I,E)$. 
Let each local control input be generated in accordance with (\ref{eq:vareps})-(\ref{eq:sam}). 
Then, for every initial condition, the state $x$
converges in a finite time to a point belonging to the set $\mathcal{E}$ in (\ref{eq:setE}).
Moreover, $\max_{i\in I}x_{i}(t)\le\overline x$ and $\min_{i\in I} x_i(t)\ge \underline x$ for 
every $t \in \mathbb R_{\geq 0}$.
\smallskip
\end{theorem}

\emph{Proof.} We start with showing the last property. We only show that 
$\max_{i \in I} x_i(t) \leq \overline x$ for every $t \in \mathbb R_{\geq 0}$ since the other case is analogous. 
We prove the claim by contradiction. 
Suppose there exists a time $t_*$ such that
$\max_{i \in I} x_i(t_*) = \overline x$ and $u_i(t_*)>0$, with 
$i$ being the index of the node exceeding $\overline x$ for the first time
(clearly, more than one node could exceed $\overline x$ at the same time
but this does not affect the analysis).
Note that $t^*$ cannot be a switching time for node $i$. 
In fact, if this were true, then we would have $u_i(t_*)>0$, which would require 
$\ave_i(t_*)\ge \eps_i (t_*)>0$, which is not possible because $x_s(t_*)\le \overline x = x_i(t_*)$ 
for all $s\in I$, by definition of $t_*$ and $i$. Thus, we focus on the case where
$t^*$ is not a switching time.

Let $t^i_k$ be the last sampling instant smaller
than $t_*$, which implies $x_s(t^i_k) \leq \overline x$ for all $s \in I$. 
Notice that $t^i_k$ is well defined even if $t_*$ occurs during the first inter-sampling interval 
of node $i$ because $x_s(0) \leq \overline x$ for all $s \in I$.
Since $u_i(t)= 1$ for all $t\in [t^i_k, t^i_{k+1}[$, it holds that
\begin{eqnarray} 
x_i(t) = x_i(t^i_k) + (t-t^i_k)
\end{eqnarray} 
Evaluating the last identity at $t=t_*$, we get 
\begin{eqnarray}  \label{eval}
\overline x - x_i(t^i_k)=t_*-t^i_k< t_{k+1}^i-t^i_k=\Delta_k^i
\end{eqnarray}
Observe now that in order for $x_i$ to grow we must also have
$|\ave_i(t_k^i)| = \ave_i(t_k^i) \ge \eps_i(t_k^i)$. 
This requires $x_i(t^i_k) < \overline x$.
In fact, if $x_i(t^i_k) = \overline x$ then node $i$ could not grow as
$x_s(t^i_k) \leq \overline x$ for all $s \in I$.
By (\ref{eq:sam}), we have
{\setlength\arraycolsep{2pt}
\begin{eqnarray}
\Delta_k^i &=& \frac{1}{4d_i} |\ave_i(t_k^i)| \nonumber \\
&=& \frac{1}{4d_i} \sum_{j \in \mathcal N_i} \left(x_j(t_k^i) - x_i(t_k^i) \right) \nonumber\\
&\leq& \frac{1}{4} ( \overline x - x_i(t_k^i) )
\end{eqnarray}}%
where the inequality comes again from the fact that 
$x_s(t^i_k) \leq \overline x$ for all $s \in I$.
The proof follows recalling the inequality \eqref{eval}.
In fact, this implies
\begin{eqnarray}
\overline x - x_i(t^i_k) < \Delta_k^i \leq \frac{1}{4} ( \overline x - x_i(t_k^i) )
\end{eqnarray}
which is not possible since $\overline x - x_i(t^i_k) \geq0$.

We now focus on the property of convergence. 
Consider the Lyapunov function
\begin{eqnarray} \label{eq:Lyap}
V(x) := \frac{1}{2}{x^T L x}
\end{eqnarray}
where $L$ is the Laplacian matrix related to the graph $G$. 
By letting $t^i_k = \max \{t^i_h \leq t, h \in \mathbb N_0\}$, 
the evolution of $V$ along the solutions to (\ref{eq:modelloA-cont}) satisfies
{\setlength\arraycolsep{2pt}
\begin{eqnarray}
\dot V(x(t)) &=& u^\top(t) L x(t) \nonumber \\
&=& -\sum_{i=1}^n \ave_{i} (t) \, \textrm{sign}_{\eps_i(t_k^i) }(\ave_{i} (t_k^i)) \nonumber \\
&=& -\sum_{i:|\ave_{i} (t_k^i)| \geq \eps_i(t_k^i)} \ave_{i} (t) \, \textrm{sign}_{\eps_i(t_k^i) }(\ave_{i} (t_k^i)) \nonumber \\
&=& -\sum_{i:|\ave_{i} (t_k^i)| \geq \eps_i(t_k^i)} \ave_{i} (t) \, \textrm{sign}(\ave_{i} (t_k^i)) 
\end{eqnarray}}%
where the last equality follows from the definition of the quantized sign function.
Observe now that if $\ave_{i} (t_k^i) \geq \eps_{i} (t_k^i)$ then
{\setlength\arraycolsep{2pt}
\begin{eqnarray}
\ave_{i} (t) &\geq& \ave_{i} (t^i_k) - 2d_i(t-t^i_k) \nonumber \\
&\geq & \ave_i(t_k^i) -\frac{1}{2}|\ave_i(t_k^i)| \nonumber\\
&= & \ave_i(t_k^i)-\frac{1}{2}\ave_i(t_k^i) \nonumber\\
&= & \frac{\ave_{i} (t^i_k)}{2}
\end{eqnarray}}%
for all $t \in [t^i_k,t^i_{k+1}]$.
This implies that $\ave_{i} (t)$ preserves  the  sign  during  continuous
flow. Similarly, if $\ave_{i} (t_k^i) \leq - \eps_{i} (t_k^i)$ then
{\setlength\arraycolsep{2pt}
\begin{eqnarray}
\ave_{i} (t) &\leq& \ave_{i} (t^i_k) + 2d_i(t-t^i_k) \nonumber \\
&\le & \frac{\ave_{i} (t^i_k)}{2}
\end{eqnarray}}%
Hence,
{\setlength\arraycolsep{2pt}
\begin{eqnarray}
\ave_{i} (t) \, \textrm{sign}(\ave_{i} (t_k^i)) &=& \ave_{i} (t) \, \textrm{sign}(\ave_{i} (t) ) \nonumber \\
&=& | \ave_{i} (t) |
\end{eqnarray}}%
This leads to
{\setlength\arraycolsep{2pt}
\begin{eqnarray}
\dot V(x(t)) 
&\leq& -\sum_{i:|\ave_{i} (t_k^i)| \geq \eps_i(t_k^i)} | \ave_{i} (t) | \nonumber \\
&\leq& -\sum_{i:|\ave_{i} (t_k^i)| \geq \eps_i(t_k^i)} \frac{| \ave_{i} (t^i_k) |}{2} \nonumber \\
&\leq& -\sum_{i:|\ave_{i} (t_k^i)| \geq \eps_i(t_k^i)} \frac{\eps}{2} 
\end{eqnarray}}%
since $\eps_i(t) \geq \eps$ for all $t \in \mathbb R_{\geq 0}$. 
Thus, there exists a finite time $T$ such that each node 
satisfies $|\ave_{i} (t_k^i)| \leq \eps_i(t_k^i)$ for every $k$ such that $t^i_k \geq T$, 
otherwise $V$ would take on negative values.
This shows that all the controls eventually become zero, which implies 
that $x(t)=x(T)$ for all $t\geq T$. Hence, we also have 
$\eps_i(t)=\eps_i(T)$ for all $t\geq T$ and for all $i \in I$.
Since the network state remains within the initial envelope, 
we have $\eps_i(t)\le \max\{\eps, \eps\chi_0\}$ for all $t \in \mathbb R_{\geq 0}$ and for all $i \in I$,
which yields the desired result. \qedp

\section{Noisy Case}
\label{sec:Noisycase}

In this section, we study convergence and boundedness properties of the proposed scheme 
in the presence of noise. 
We first show that the proposed coordination method ensures 
boundedness of the state trajectories. 

\subsection{Boundedness of the state trajectories}

Let
\begin{eqnarray}
 \gamma:=\left(\frac{1}{3}+\frac{4}{3} \frac{d_{max}}{\eps}\right)|w|_\infty
 \label{eq:gamma}
\end{eqnarray}
We have the following result. \smallskip
 
 \begin{theorem}\label{thm:statebounded}
	Consider a network of $n$ dynamical systems as in (\ref{eq:modelloA-cont}), which are
	interconnected over an undirected connected graph $G = (I,E)$. 
	Let each local control input be generated in 
	accordance with (\ref{eq:vareps})-(\ref{eq:sam}). Then, for every initial condition, the state $x$ satisfies
	\begin{eqnarray}\label{eq:xsup}
	\max_{i\in I} x_i(t) \leq 
	\left\{
	\def\arraystretch{1.2}
	\begin{array}{rl}
	\overline x & \mathrm{if}\ |\overline x|\ge \gamma\\ 
	\gamma & \mathrm{otherwise}
	\end{array} \right.
	\end{eqnarray}
	and
	\begin{eqnarray}\label{eq:xinf}
	\min_{i\in I} x_i(t) \geq 
	\left\{
	\def\arraystretch{1.2}
	\begin{array}{rl}
	\underline x & \mathrm{if}\ |\underline x|\ge \gamma\\ 
	-\gamma & \mathrm{otherwise}
	\end{array} \right.
	\end{eqnarray}
	for 
every $t \in \mathbb R_{\geq 0}$.
\end{theorem} \smallskip

\emph{Proof.}  We will only prove the result regarding $\max_{i \in I} x_i(t)$ 
since the other can be proved in an analogous manner. Notice that $\ave^w_i(t) =  \ave_i(t) + \phi_i(t)$
for all $t \in \mathbb R_{\geq 0}$ and all $i \in I$, where we defined
\begin{eqnarray} 
\phi_i(t) := \sum_{j \in \mathcal N_i} w_j(t)
\end{eqnarray} 
Clearly, we have 
\begin{eqnarray} \label{eq:phibound}
| \phi_i(t) | \leq d_{max}|w|_\infty
\end{eqnarray}
for all $t \in \mathbb R_{\geq 0}$ and all $i \in I$.
 
\emph{Case 1: $|\overline x| \geq \gamma$.}
We show that there is no node {that} can exceed $\overline x$. 
Suppose that there exists a time $t_*$ such that
$\max_{i \in I} x_i(t_*) = \overline x$ and $u_i(t_*)>0$, with 
$i$ the index of the  node 
exceeding $\overline x$ {for the first time} (clearly, more than one node could exceed $\overline x$ at the same time
but this does not affect the analysis). In contrast with the proof of 
Theorem \ref{thm:noiseless}, here $t_*$ may potentially be a switching time, since it could happen that 
$\ave_i^w (t_*)\ge \eps_i (t_*)$ even though $x_s(t_*)\le \overline x = x_i(t_*)$ for all $s\in I$ 
due to the presence of the noise $w$. 
The case in which $t_*$ is a switching instant falls into the case studied in the next paragraph.

Let $t^i_k$ be the last sampling instant not greater than $t_*$, 
which implies $x_s(t^i_k) \leq \overline x$ for all $s \in I$.
Notice that $t^i_k$ is well defined even if $t_*$ occurs in the first inter-sampling interval 
of node $i$ since $x_s(0) \leq \overline x$ for all $s \in I$.
 We have two sub-cases.

 \emph{Sub-case 1: $x_i(t^i_k) > \overline x - \frac{1}{3} |w|_\infty$.}
The condition for $x_i$ to grow is 
\begin{eqnarray} 
\ave^w_i(t^i_k) =  \ave_i(t^i_k) + \phi_i(t^i_k) \geq \eps_i(t^i_k)
\end{eqnarray} 
Since $x_s(t^i_k) \leq \overline x$ for all $s \in I$, we have
{\setlength\arraycolsep{2pt}
\begin{eqnarray} \label{eq:avephibound}
\ave_i(t^i_k) &\leq& d_i \overline x - d_i x_i(t^i_k) \nonumber \\
&\leq& d_i(\overline x-(\overline x-\frac{1}{3}|w|_\infty)) \nonumber \\
&\leq& \frac{1}{3} d_{max} |w|_\infty
\end{eqnarray}}%
By combining (\ref{eq:phibound}) and (\ref{eq:avephibound}), in order for 
$x_i$ to grow we must necessarily have
\begin{eqnarray} 
\frac{4}{3} d_{max} |w|_\infty \geq \eps_i(t^i_k) 
\end{eqnarray} 
This leads to a contradiction. In fact, if $|x_i(t^i_k)| \geq 1$ then $\eps_i(t^i_k) = \eps |x_i(t^i_k)|$. 
Moreover,
$|x_i(t^i_k)| > |\overline x| - \frac{1}{3} |w|_\infty$.
Hence, we must necessarily have
\begin{eqnarray} 
\frac{4}{3} d_{max} |w|_\infty 
> \eps ( |\overline x| - \frac{1}{3} |w|_\infty )
\end{eqnarray} 
which implies $|\overline x| < \gamma$, thus leading to a contradiction. 
If instead 
$|x_i(t^i_k)| < 1$ then $\eps_i(t^i_k) = \eps$ and we must have 
\begin{eqnarray} 
\frac{4}{3} d_{max} |w|_\infty \geq \eps 
\end{eqnarray} 
This leads again to a contradiction since, by hypothesis, we must have $\gamma \leq |\overline x|$
and $|\overline x| < |x_i(t^i_k)| +  \frac{1}{3} |w|_\infty < 1 + \frac{1}{3} |w|_\infty$. This would imply 
$\frac{4}{3} d_{max} |w|_\infty < \eps$.

 \emph{Sub-case 2: $x_i(t^i_k) \leq \overline x - \frac{1}{3} |w|_\infty$.}
By construction, $x_i$ can grow at most up to
\begin{eqnarray} 
&& x_i(t^i_k) + \frac{1}{4d_i} ( \ave_i(t^i_k) + \phi_i(t^i_k) )  \nonumber \\
&& \qquad = \frac{3}{4} x_i(t^i_k) + \frac{1}{4d_i}  
\sum_{j \in \mathcal N_i} (x_j(t^i_k) + w_j(t^i_k) ) \nonumber \\
&& \qquad \leq \frac{3}{4} x_i(t^i_k) + \frac{1}{4} ( \overline x + |w|_\infty  ) \label{grow.up.to}
\end{eqnarray} 
where the inequality follows 
since $x_s(t^i_k) \leq \overline x$ for all $s \in I$.
Since $x_i(t^i_k) \leq \overline x - \frac{1}{3} |w|_\infty$
we conclude that $x_i$ can grow at most up to
\begin{eqnarray} 
\frac{3}{4} (\overline x - \frac{1}{3} |w|_\infty) + \frac{1}{4} ( \overline x + |w|_\infty  ) = \overline x \label{grow.up.to2}
\end{eqnarray}  
which leads to a contradiction.

\emph{Case 2. $|\overline x| < \gamma$.}
The proof of this case is exactly same as for the previous case with $\overline x$ replaced by $\gamma$. \qedp

\subsection{Consensus properties under low-magnitude noise}

We start with a simple result which shows that convergence 
is preserved under noise  whenever $|w|_\infty$ is sufficiently small compared to $\eps$. 
Moreover, the state remains within the initial envelope like in the noiseless case. \smallskip
 
\begin{theorem}\label{thm:smallnoise}
Consider a network of $n$ dynamical systems as in (\ref{eq:modelloA-cont}), which are
interconnected over an undirected connected graph $G = (I,E)$. 
Let each local control input be generated in accordance with (\ref{eq:vareps})-(\ref{eq:sam}). 
Suppose that $\eps>2d_{max}|w|_\infty$. Then, for every initial condition, the state $x$
converges in a finite time to a point belonging to the set $\mathcal{D}$ in (\ref{eq:setD}).
Moreover, $\max_{i\in I}x_{i}(t)\le\overline x$ and $\min_{i\in I} x_i(t)\ge \underline x$ for all $t \in \mathbb R_{\geq 0}$.
\end{theorem} \smallskip
 
\emph{Proof.} We first show the last property. 
This can be done following the same steps as {in} the noiseless case.
Again, we only show that $\max_{i \in I} x_i(t) \leq \overline x$
for all $t \in \mathbb R_{\geq 0}$.
Suppose that there exists a time $t_*$ such that
$\max_{i \in I} x_i(t_*) = \overline x$ and $u_i(t_*)>0$, with 
$i$ the index of the first node 
exceeding $\overline x$ (clearly, more than one node could exceed $\overline x$ at the same time
but this does not affect the analysis). 
Let $t^i_k$ be the last sampling instant not greater than $t_*$, which implies $x_s(t^i_k) \leq \overline x$ for all $s \in I$. 
Notice that $t^i_k$ is well defined even if $t_*$ occurs during the first inter-sampling interval 
of node $i$ because $x_s(0) \leq \overline x$ for all $s \in I$.
Clearly, we must necessarily have $|\ave^w_i(t_k^i)| = \ave^w_i(t_k^i) \ge \eps_i(t_k^i)$.
Moreover,
\begin{eqnarray}\label{eq:bo}
x_i(t) \leq x_i(t^i_k) + (t-t^i_k)
\end{eqnarray}
for all $t \in [t^i_k,t^i_{k+1}]$.

By (\ref{eq:sam}), we have
{\setlength\arraycolsep{2pt}
\begin{eqnarray}
\Delta^i_k &=& \frac{1}{4d_i} |\ave_i^w(t_k^i)| \nonumber \\
&=& \frac{1}{4d_i}  \sum_{j \in \mathcal N_i} \left(x_j(t_k^i) - x_i(t_k^i) + w_j(t_k^i)\right ) \nonumber\\
&\leq& \frac{1}{4} ( \overline x - x_i(t_k^i) +|w|_\infty )
\end{eqnarray}}%
where the inequality follows from the fact that $x_s(t^i_k) \leq \overline x$ for all $s \in I$.
By hypothesis, $t^i_k$ is the last sampling instant not greater than $t_*$. Hence, 
since the control input is constant over $[t^i_k,t^i_{k+1}]$ and because $x_i$ must exceed  
$\overline x$ we must have $\overline x < x_i(t_{k+1}^i)$.
Hence,
\begin{eqnarray}
\overline x - x_i(t^i_k) < \Delta^i_k \leq \frac{1}{4} ( \overline x - x_i(t_k^i)+ |w|_\infty )
\end{eqnarray}
This inequality is possible only when
\begin{eqnarray}
\overline x-x_i(t_k^i)<\frac{1}{3}|w|_\infty
\end{eqnarray}
However, this implies
{\setlength\arraycolsep{2pt}
\begin{eqnarray}
\ave_i^w(t_k^i) &=& \sum_{j \in \mathcal N_i} \left(x_j(t_k^i) - x_i(t_k^i) + w_j(t_k^i)\right)\nonumber \\
&\leq& d_{max}(\overline x -x_i(t_k^i) +|w|_\infty) \nonumber\\
&< & \frac{4}{3} d_{max} |w|_\infty  \nonumber\\
&< & \eps
\end{eqnarray}}%
where the last inequality follows since $2d_{max}|w|_\infty < \eps$ by hypothesis. 
This implies that $\ave_i^w(t_k^i) < \eps_i(t_k^i)$, thus leading to a contradiction.

We now focus on convergence. Let $V$ be defined as in (\ref{eq:Lyap}), 
and consider the evolution of $V$ along the solutions to (\ref{eq:modelloA-cont}). 
By letting $t^i_k = \max \{t^i_h \leq t, h \in \mathbb N_0\}$, we have
{\setlength\arraycolsep{2pt}
\begin{eqnarray}
\dot V(x(t)) &=& u^\top(t) L x(t) \nonumber \\
&=& -\sum_{i=1}^n \ave_{i} (t) \, \textrm{sign}_{\eps_i(t_k^i) }(\ave^w_{i} (t_k^i)) \nonumber \\
&=& -\sum_{i:|\ave^w_{i} (t_k^i)| \geq \eps_i(t_k^i)} \ave_{i} (t) \, 
\textrm{sign}_{\eps_i(t_k^i) }(\ave^w_{i} (t_k^i)) \nonumber \\
&=& -\sum_{i:|\ave^w_{i} (t_k^i)| \geq \eps_i(t_k^i)} \ave_{i} (t) \, \textrm{sign}(\ave^w_{i} (t_k^i)) 
\end{eqnarray}}%
where the last equality follows from the definition of the quantized sign function.
Observe now that if $\ave^w_{i} (t_k^i) \geq \eps_{i} (t_k^i)$ 
then $\textrm{sign}(\ave^w_{i} (t_k^i)) =1$.
Moreover,
{\setlength\arraycolsep{2pt}
\begin{eqnarray} \label{eq:avegeeps}
\ave_{i} (t) &\geq& \ave_{i} (t^i_k) - 2d_i(t-t^i_k) \nonumber \\
&\geq& \ave_{i} (t^i_k) - \frac{1}{2} | \ave^w_{i} (t^i_k) |  \nonumber \\
&\geq& \ave_{i} (t^i_k) - \frac{1}{2} \ave^w_{i} (t_k^i) \nonumber \\
&=& \frac{1}{2}  \ave^w_{i} (t^i_k) - \phi_{i} (t^i_k) \nonumber \\
&\geq& \frac{1}{2}  \varepsilon - d_{max} |w|_\infty
\end{eqnarray}}%
for all $t \in [t^i_k,t^i_{k+1}]$. Similarly, if $\ave^w_{i} (t_k^i) \leq - \eps_{i} (t_k^i)$ then
$\textrm{sign}(\ave^w_{i} (t_k^i)) =-1$, and 
{\setlength\arraycolsep{2pt}
\begin{eqnarray}
\ave_{i} (t) &\leq& \ave_{i} (t^i_k) + 2d_i(t-t^i_k) \nonumber \\
&\leq& \ave_{i} (t^i_k) + \frac{1}{2} | \ave^w_{i} (t^i_k) |  \nonumber \\
&\leq& - \frac{1}{2}  \varepsilon + d_{max} |w|_\infty
\end{eqnarray}}%
This leads to
\begin{eqnarray}
\dot V(x(t)) 
\leq -\sum_{i:|\ave^w_{i} (t_k^i)| \geq \eps_i(t_k^i)} \left( \frac{1}{2}  \varepsilon - d_{max} |w|_\infty \right) 
\end{eqnarray}
for all $t \geq 0$. Since $\varepsilon > 2 d_{max} |w|_\infty$,
\begin{eqnarray}
\frac{1}{2}  \varepsilon - d_{max} |w|_\infty = \alpha 
\end{eqnarray}
for some $\alpha>0${, since} all the quantities involved are constant.
Hence, there exists a finite time $T'$ after which each node 
satisfies $|\ave^w_{i} (t_k^i)| < \eps_i(t_k^i)$ for every $k$ such that $t^i_k \geq T'$,
otherwise $V$ would take on negative values.
Since $x$ remains within the initial envelope then
$|\ave^w_{i} (t)| \leq d_i (2 \chi_0 +  |w|_\infty)$
for all $t \in \mathbb R_{\geq 0}$. Thus
$\Delta^i_k \leq \max\{\eps, (2 \chi_0 +  |w|_\infty)\}/4 := \bar \Delta$ 
for every $k \in \mathbb N_0$.
This shows that all the controls eventually become zero not later than $T:=T'+\bar \Delta$, 
which implies that $x_i(t)=x_i(T)$ and $\ave_i(t)=\ave_i(T)$ for all $t\geq T$. 
Moreover, since $x$ remains within the initial envelope
we also have $\eps_i(t) \leq \max\{\eps, \eps\chi_0\}$ for all $t \in \mathbb R_{\geq 0}$.
Taking any $t^i_k \geq T$ we then have
{\setlength\arraycolsep{2pt}
\begin{eqnarray} \label{eq:smnoisresult}
|\ave_i(t)| &=& |\ave_i(t^i_k)| \nonumber \\
&\leq& |\ave^{w}_{i} (t_k^i)| + d_{max}  |w|_\infty  \nonumber \\
&\leq& \max\{\eps, \eps\chi_0\} + d_{max}  |w|_\infty 
\end{eqnarray}}% 
The proof is concluded by noting that the right side of (\ref{eq:smnoisresult}) 
is upper bounded by $r$. \qedp

\subsection
{Consensus properties under general noise}

In general, condition $\eps > 2 d_{max} |w|_\infty$ need not be satisfied 
if $|w|_\infty$ is unknown. Even if $|w|_\infty$ is known, enforcing this condition  
might lead to large errors between network nodes.  
To this end, we study the properties of the proposed approach
for the general {case of} noise  which are unknown but bounded. 
We have the following result. \smallskip

\begin{theorem} \label{thm:generalnoise}
Consider a network of $n$ dynamical systems as in (\ref{eq:modelloA-cont}), 
which are
interconnected over an undirected connected graph $G = (I,E)$. 
Let each local control input be generated in accordance with (\ref{eq:vareps})-(\ref{eq:sam}). 
Then, for every initial condition, the network state $x$ 
enters in a finite time the set $\mathcal{D}$ in (\ref{eq:setD}) and remains there forever. 
Moreover, $x$ converges {in a finite time to a point belonging to the set 
$\mathcal{D}$ in (\ref{eq:setD})} when the noise  converge to zero.
\end{theorem} \smallskip
  
We prove two technical results which are instrumental for the 
proof of Theorem \ref{thm:generalnoise}. 

The first result relates $\eps_i$ and $L$. \smallskip

\begin{lemma}\label{lem:elessL}
Consider the same assumptions and conditions as in Theorem \ref{thm:generalnoise}. 
For any $i\in I$, it holds that
\begin{eqnarray}
\eps_i(t_k^i) \leq  r-\frac{5}{3}d_{max}|w|_{\infty}
\end{eqnarray}
for every $k \in\mathbb{N}_0$. \smallskip
\end{lemma}

\emph{Proof.} By Theorem \ref{thm:statebounded}, we have 
 \begin{eqnarray}
  |x_i(t_k^i)| \leq \max \{|\overline x|,|\underline x|,\gamma\}\le \chi_0+\gamma
 \end{eqnarray}
 Hence,
{\setlength\arraycolsep{2pt}
 \begin{eqnarray}
 \eps_i(t_k^i)&=&\max\{\eps,\eps|x_{i}(t_k^i)|\}\nonumber\\
 &\leq& \max\{\eps,\eps(\chi_0+\gamma)\}\nonumber\\
 &\leq& \max\{\eps,\eps\chi_0\}+\eps \gamma \nonumber\\
 &=& r-\frac{5}{3}d_{max}|w|_\infty \label{eq:lem1}
 \end{eqnarray}}%
where the last equality holds by the definitions \eqref{eq:r} 
and \eqref{eq:gamma} of $r$ and $\gamma$ respectively.\qedp 
  
The second result shows that the average preserves the sign as 
long as its absolute value remains large enough compared with the radius $r$. 
\smallskip

 \begin{lemma} \label{lem:samesign} 
 	Consider the same assumptions and conditions as in Theorem \ref{thm:generalnoise}.
 	Consider any index $i \in I$ and any 
	$M \in \mathbb N_0$. 
	If $|\ave_{i} (t^i_{k+m})| \ge r$ for $m = 0,1,\ldots,M$
	then
 	\begin{eqnarray}
    \sign(\ave_{i} (t^i_{k+m})) = \sign(\ave_{i} (t^i_{k})), \nonumber \\
    m = 1,2,\ldots,M+1
 	\end{eqnarray}
 \end{lemma} 
 \emph{Proof.}  Assume without loss of generality that $\ave_{i} (t^i_{k})\ge r$,
 the other case being analogous. From Lemma \ref{lem:elessL}, we have
 {\setlength\arraycolsep{2pt}
 \begin{eqnarray}
 \ave_{i}^w(t_k^i)& \geq &\ave_{i} (t^i_{k})-d_{max}|w|_\infty \nonumber\\
 & \geq & r-d_{max}|w|_\infty \nonumber\\
 & \geq & \eps_i(t_k^i)
\end{eqnarray}}%
Hence, $u_{i} (t^i_{k}) = 1$. Moreover,
{\setlength\arraycolsep{2pt}
\begin{eqnarray} \label{}
\ave_{i} (t) &\geq& \ave_{i} (t^i_k) - 2d_i(t-t^i_k) \nonumber \\
&\geq& \ave_{i} (t^i_k) - \frac{1}{2} \ave^w_{i} (t_k^i) \nonumber \\
&=& \frac{1}{2}  \ave_{i} (t^i_k) - \frac{1}{2}  \phi_{i} (t^i_k) \nonumber \\
&\geq& \frac{1}{2} r - \frac{1}{2} d_{max} |w|_\infty \nonumber \\
&\geq& \frac{1}{2} \max \{\eps,\eps \chi_0\}
\end{eqnarray}}% 
for all $t \in [t^i_k,t^i_{k+1}]$.

We then conclude that $\ave_{i} (t^i_{k+1}) >0$.
Thus $\ave_{i}$ preserves its sign. \qedp \smallskip
 
We can now proceed with the proof of Theorem \ref{thm:generalnoise}. \smallskip
 
\emph{Proof of Theorem \ref{thm:generalnoise}.} 
We only show the result for the case $\eps\le 2d_{max}|w|_\infty$ 
since the other case can be derived from Theorem \ref{thm:smallnoise}.
To begin with, we 
introduce three sets into which we partition the set of switching times of each node $i$.
For each $i \in I$, let 
{\setlength\arraycolsep{1pt} 
\begin{eqnarray}
\def\arraystretch{1.5}
\begin{array}{l}
\mathscr S_{i1} := \big\{ t^i_k  : \,  |\ave^w_{i} (t^i_k)| \geq \eps_{i} (t^i_k)   \wedge  
 |\ave_{i} (t^i_k)| \geq r \big\}   \\ 
\mathscr S_{i2} := \big\{ t^i_k  :  \, |\ave^w_{i} (t^i_k)| \geq \eps_{i} (t^i_k) \wedge
|\ave_{i} (t^i_k)| < r  \big\}  \\
\mathscr S_{i3} := \left\{ t^i_k  : \,  |\ave^w_{i} (t^i_k)|
< \varepsilon_{i} (t^i_k) \right\} 
\end{array}  
\end{eqnarray}
Clearly, $t^i_k \in \mathscr S_{i1} \cup \mathscr S_{i2} \cup \mathscr S_{i3}$ 
for every $k \in \mathbb N_0$.
 
Pick any $i \in I$, and assume by contradiction that there exists a time $t_*$ such that
$|\ave_{i} (t^i_{k})| \geq r$ for all $t^i_{k}\geq t_*$. 
In view of Lemma \ref{lem:elessL}, $u_i$ is
never zero from $t_*$ on 
since the condition above yields 
$|\ave^w_{i} (t^i_{k})| \geq r - d_{max} |w|_\infty \geq \eps_i(t^i_k)$.
Moreover, by Lemma \ref{lem:samesign}, 
$\sign(\ave_{i} (t^i_{k+m})) = \sign(\ave_{i} (t^i_k))$ for every $m$. 
Hence, either $u_i(t) = 1$ for all $t^i_{k} \geq t_*$ 
or $u_i = -1$ for all $t^i_{k} \geq t_*$. This would imply that $x_i$ diverges,
violating the state boundedness property of Theorem \ref{thm:statebounded}.
  
 By the foregoing arguments, there exists a time instant $t^i_k$ such that 
 $|\ave_{i} (t^i_{k})| <  r$. This implies that $t^i_{k} \notin \mathscr S_{i1}$, {or, equivalently, 
 that $t^i_{k} \in \mathscr S_{i2}\cup \mathscr S_{i3}$.} 
 Thus it remains to show that transitions from $\mathscr S_{i2}$ and $\mathscr S_{i3}$ 
 to $\mathscr S_{i1}$ are not possible. We analyze the two cases separately.

\emph{Case 1: $t^i_{k} \in \mathscr S_{i2}$.}
In this case, $u_i(t^i_k) = \{-1,1\}$. 
Suppose that $u_i(t^i_k)=1$, 
the other case being analogous. Then, 
{\setlength\arraycolsep{2pt} 
\begin{eqnarray}
\ave_{i} (t) \leq \ave_{i} (t^i_k) < r 
\end{eqnarray}}%
for all $t \in [t^i_k,t^i_{k+1}]$ 
where the first inequality follows since $u_i(t^i_k)=1$
while the second inequality follows because $t^i_k \in \mathscr S_{i2}$ by hypothesis.
In addition, condition $u_i(t^i_k)=1$
implies $\ave_{i} (t^i_k) \geq \varepsilon_{i} (t^i_k) - \phi_{i} (t^i_k)$.
Thus, 
{\setlength\arraycolsep{2pt} 
\begin{eqnarray} \label{eq:pointwise_invariance_1}
\ave_{i} (t) &\geq& \ave_{i} (t^i_k) - 2d_i (t-t^i_k)  \nonumber \\
&=& \ave_{i} (t^i_k) -  \frac{1}{2} \ave^w_{i} (t^i_k)  \nonumber \\
&\geq& \frac{1}{2}   \ave_{i} (t^i_k) -  \frac{1}{2} d_{max} |w|_\infty   \nonumber \\
&\geq& \frac{1}{2}  \varepsilon_{i} (t^i_k) - d_{max} |w|_\infty  \nonumber \\
&> &  - d_{max} |w|_\infty  \nonumber \\
&>& - r  
\end{eqnarray}}%
for all $t \in [t^i_k,t^i_{k+1}]$. 
Thus $|\ave_{i} (t^i_{k+1})| < r$ which implies that $t^i_{k+1} \notin \mathscr S_{i1}$.
 
\emph{Case 2: $t^i_{k} \in \mathscr S_{i3}$.}
In this case we have
${u_i}(t)=0$ for all $t\in[t_k^i, t_{k+1}^i]$ and $t^i_{k+1}-t^i_k= \varepsilon/(4d_i)$. Hence,
{\setlength\arraycolsep{2pt} 
 \begin{eqnarray}\label{eq:avethm4p}
 |\ave_{i} (t)| &\leq& |\ave_{i} (t^i_k)| + d_i (t-t^i_k)  \nonumber \\
 &<& \eps_{i} (t^i_k) + d_{max}|w|_\infty +\frac{\eps}{4}  \nonumber \\
 &<& \eps_{i} (t^i_k) +\frac{3}{2} d_{max}|w|_\infty \nonumber \\
 &<& r
 \end{eqnarray}}%
{for all $t \in [t^i_k,t^i_{k+1}]$,}
where the {third} inequality 
follows from $\eps\le 2d_{max}|w|_\infty$ and 
the fourth one follows from Lemma \ref{lem:elessL}. Hence, $t^i_{k+1} \notin \mathscr S_{i1}$. 

Hence, we conclude that 
$t^i_\ell\in \mathscr S_{i2} \cup \mathscr S_{i3}$ for all $\ell\ge k$. Moreover, the previous arguments 
show that $ |\ave_{i} (t)|<r$ for all $t\in [t^i_\ell,t^i_{\ell+1}]$, for all $\ell\ge k$, 
which guarantees that $x$
remains forever inside $\mathcal D$. Finally, if $w$ converges to zero then there exists a finite instant $t_*$
 such that $\eps > 2 d_{max} \sup_{t \geq t_*}|w(t)|$, and the convergence result
 follows along the same lines as in Theorem \ref{thm:smallnoise}.
 \qedp \smallskip
 
\begin{remark} \label{rem:nonstatic}
  	In contrast with the noiseless case (Theorem \ref{thm:noiseless})
	and the case of low-magnitude noise (Theorem \ref{thm:smallnoise}), 
	one sees that in the general case the network nodes need not converge
	but remain confined in a neighbourhood of consensus that depends 
	on both $\eps$ and $w$. \qedp
  \end{remark}   

%%%%%%%%%%%%%%%%%%%%%%%%%%%%%%%%%%%
%%%%%%%%%%%%%%%%%%%%%%%%%%%%%%%%%%%
 
\section{Adaptive Thresholds, Sign Function and
Node-to-node Error} \label{sec:n2n_error}

In this section, we further comment on the considered notion of consensus
and discuss a number of properties ensured by the proposed coordination scheme. 

\subsection{Adaptive thresholds and sign function}

The main problem when dealing with communication noise is that the Laplacian graph
matrix has an eigenvalue in zero. This may cause the state to drift when 
the noise has non-zero mean. In this paper, drifting is prevented 
by resorting to local adaptive thresholds
\begin{eqnarray}\label{}
\eps_i(t) := 
\def\arraystretch{1.2}
\left\{ \begin{array}{ll}
\eps |x_i(t)| & \quad \text{if}\ |x_i(t)| \geq 1 \\ 
\eps & \quad \text{otherwise}
\end{array} \right.
\end{eqnarray} 
These adaptive thresholds scale with the magnitude of the data
and this feature is essential to guarantee that any drifting will eventually 
stop. Specifically, recall that the local control action is given by
\begin{eqnarray} 
u_i(t) = \sign_{\eps_i(t^i_k)}\!\left(\ave^w_i(t^i_k)\right) 
\end{eqnarray}
where 
\begin{eqnarray} \label{eq:ave_noisy}
\ave^w_i(t) =  \ave_i(t) + \sum_{j \in \neigh{i}} w_j(t)
\end{eqnarray}
Suppose that $x_i$ starts drifting, for example growing ($u_i \equiv 1$). 
Since $u_i \equiv 1$ then $\ave_i = \sum_{j \in \neigh{i}} (x_j-x_i)$ cannot grow, so that $\ave^w_i$ 
must remain bounded. Hence, 
adapting the threshold of the $\sign$ function  to the magnitude of $x_i$
eventually forces 
$\eps_i$ to become larger than $\ave^w_i$. We will exemplify this feature in Section \ref{Exm:smallgraph}. In contrast,  a pure constant $\eps$ need not counteract the drifting of $x_i$ since $\ave^w_i$ 
may persistently remain larger than $\eps$. 

Another interesting feature of the proposed scheme lies in the use of the
$\sign$ function. When the level of disagreement 
is large compared with the noise magnitude, 
for example during the initial phase of coordination, then $\ave^w_i\approx\ave_i$. 
In this situation, the $\sign$ function ensures that the control action 
will be the same as in the noiseless case.
In other terms, the noise will affect coordination
only when nodes are sufficiently close to consensus.
Also this feature will be exemplified in Section VII-A.

The $\sign$ function does also permit to save communication resources, 
which is one of the main issues when coordination is carried out through packet-based networks.
Recall that in the proposed scheme the inter-transmission times $\Delta^i_k$ are defined as
\begin{eqnarray} \label{}
\Delta^i_k:=
\def\arraystretch{2.2}
\left\{
\begin{array}{ll}
\displaystyle \frac{|\ave^w_i(t^i_{k})|}{4d_{i}} & \quad  \textrm{if } \, 
|\ave^w_i(t^i_{k})|\ge \eps_i(t^i_{k})  \\
\quad\ \displaystyle \frac{\eps}{4d_{i}} & \quad
\textrm{otherwise} 
\end{array} \right.
\end{eqnarray}
As noted before, when $\ave_i$ is large compared with the noise magnitude,
then $\ave^w_i\approx\ave_i$ and the control action behaves as in the noiseless case.
In the proposed scheme, condition $\ave^w_i\approx\ave_i$ is implemented as $|\ave^w_i|\ge \eps_i$.
In particular, when $|\ave^w_i|\ge \eps_i$ then 
$\Delta^i_k$ increases with ${\ave}^w_i$ with the idea that large values of ${\ave}^w_i$
correspond to a situation where the disagreement is large so that there is no need for very frequent 
control variations. The situation is different when $|\ave^w_i| < \eps_i$.
In this case, it may happen that $\ave^w_i$ is significantly different from $\ave_i$.
Moreover, $|\ave^w_i| < \eps_i$ also implies that the level of disagreement 
is small compared with the data magnitude. Thus, if $|\ave^w_i| < \eps_i$
then $\Delta^i_k$ is decreased to $\eps/(4d_{i})$ with the idea 
that control variations should be made more frequent so as to counteract the effect of noise and 
maintain a small level of disagreement. Clearly, in this situation $\Delta^i_k$ may become
small if $\eps$ is chosen small, and the latter is desired to ensure a small level of disagreement.
As discussed in the next subsection, there is actually no need to pick $\eps$ very 
small in order to secure a small level of disagreement, which means that 
communications need not be frequent even when the nodes are within 
the consensus region.

\subsection{\textcolor{black}{Node-to-node error}}

The proposed coordination scheme guarantees that, in the noiseless case, 
all the nodes remain
between the minimum and the maximum of their initial values,
and converge in a finite time to a point belonging to the set
\begin{equation}
{\cal E} = \big\{x\in \mathbb{R}^n: 
|\sum_{j\in {\cal {N}}_i}(x_j-x_i)| < \max\{\eps, \eps \chi_{0} \},\ \forall i\in I \big\}
\end{equation}
where $\eps \in (0,1)$ is a design parameter, and $\chi_0 = |x_i(0)|_\infty$.
As noted, when $\chi_0>1$ the coordination scheme guarantees 
that, in a finite time, 
\begin{eqnarray}  \label{eq:local_average}
\frac{|\sum_{j\in {\cal {N}}_i}(x_j-x_i)|}{\chi_0} \leq \eps \quad \forall i \in I 
\end{eqnarray}
The parameter $\eps$ determines the desired accuracy level for the consensus 
final value, which is normalized to the magnitude of the initial data.  
In this way, a maximum error $\eps$ is guaranteed for the \emph{worst case} 
over the initial vector of measurements.
If instead $\chi_0\le 1$ then the tolerance becomes $\eps$.  
The parameter $\eps$ plays a crucial role for consensus.
On one side, it is desirable to choose $\eps \ll 1$ so as 
to guarantee a small level of disagreement. On the other hand,
a very small value of $\eps$ can render the coordination scheme 
very sensitive to noise. Moreover, as noted before, small values of $\eps$
can induce large communication rates since $\eps$ determines 
the smallest inter-transmission time of each node. 
It is the term $|\sum_{j\in {\cal {N}}_i}(x_j-x_i)|$ that somehow
makes this tradeoff less critical. 

At first glance, it seems indeed more natural to search for coordination
schemes that guarantee 
\begin{eqnarray} \label{eq:n2n}
\frac{|x_j-x_i|}{\chi_0} \leq \eps \quad \forall i,j \in I
\end{eqnarray}
or \emph{node-to-node} error. 
In fact, the latter guarantees that the disagreement 
is small for every pair of nodes (not necessarily connected), while (\ref{eq:local_average})
only ensures that the disagreement is small locally (for its neighbourhood).
Actually, in many cases of practical interest 
it turns out that a bound $r$
on the local averages implies a bound on the node-to-node error which is strictly smaller than $r$.
In this situation, working with (\ref{eq:local_average}) is advantageous compared with (\ref{eq:n2n})
since this guarantees a small node-to-node error without requiring to choose $\eps$ 
too small. In turn, this moderates the noise sensitivity and the number of 
communications. As discussed next, this situation happens when the network connectivity is sufficiently large.
We make this argument precise.

Consider the same setting as in Theorem \ref{thm:generalnoise}, and let $T$ denote the time 
after which the network state remains confined in $\mathcal D$. Pick any fixed time instant 
$t \geq T$ and let $x_M$ and $x_m$ denote the network nodes taking on
maximum and minimum value, respectively. The indices $M$ and $m$
may change with time but we consider a fixed $t$.
Let $\alpha:= x_M(t)-x_m(t)$ with $\alpha >0$ (the case $\alpha=0$ is not 
interesting because the network would be at perfect consensus).
By Theorem \ref{thm:generalnoise}, $|\ave_i(t)| < r$ 
for all $i \in I$. We now relate $\alpha$ and $r$.
First notice that
{\setlength\arraycolsep{1pt}
	\begin{eqnarray}
	\ave_M &=& \sum_{j \in \mathcal N_M} (x_j-x_M) \nonumber \\
	&=& d_M (x_m-x_M) + \sum_{j \in \mathcal N_M} (x_j-x_m) \nonumber \\
	&=& - d_M \alpha + \sum_{j \in \mathcal N_M} (x_j-x_m) 
	\end{eqnarray}}% 
where we omitted the time argument for brevity. Decompose
$\mathcal N_M = (\mathcal N_M \setminus \mathcal N_m) \cup \mathcal (\mathcal N_M \cap \mathcal N_m)$.
Since $x_j-x_m \leq \alpha$ for all $j \in I$, we obtain
{\setlength\arraycolsep{1pt}
	\begin{eqnarray}
	\sum_{j \in (\mathcal N_M \setminus \mathcal N_m)} (x_j-x_m) \leq \delta \alpha
	\end{eqnarray}}% 
where
{\setlength\arraycolsep{1pt}
\def\arraystretch{1.5}
	\begin{eqnarray}
	&& \delta := 
	\left\{ 
	\begin{array}{ll} 
	|\mathcal N_M \setminus \mathcal N_m| -1 & \quad \textrm{if } m \in \mathcal N_M \\ 
	|\mathcal N_M \setminus \mathcal N_m| & \quad \textrm{otherwise} 
	\end{array}
	\right. \nonumber \\
	\end{eqnarray}}% 
Moreover,
{\setlength\arraycolsep{1pt}
\begin{eqnarray}
&& \sum_{j \in (\mathcal N_M \cap \mathcal N_m)} (x_j-x_m) < \mu
\end{eqnarray}}% 
where
{\setlength\arraycolsep{1pt}
\def\arraystretch{1.5}
	\begin{eqnarray}
	&& \mu := 
	\left\{ 
	\begin{array}{ll} 
	r  - \alpha & \quad \textrm{if } M \in \mathcal N_m \\ 
	r & \quad \textrm{otherwise} 
	\end{array}
	\right. 
	\end{eqnarray}}% 
In fact, $\sum_{j \in Q} (x_j-x_m) < r$ for every set $Q \subseteq \mathcal N_m$
because $|\ave_m|<r$ and $m$ is the node that takes on the minimum value in the network. 
In addition, if $M \in \mathcal N_m$ we then have
$(\mathcal N_M \cap \mathcal N_m) \subseteq (\mathcal N_m \setminus \{M\})$, which
implies $\mu = r- \alpha$.
Since $|\ave_M|<r$, we get
{\setlength\arraycolsep{1pt}
	\begin{eqnarray}
	-r  < {\ave}_M &=& - d_M \alpha + \sum_{j \in \mathcal N_M} (x_j-x_m) \nonumber \\
	&<& - (d_M - \delta) \alpha + \mu
	\end{eqnarray}}% 
which implies
{\setlength\arraycolsep{1pt}
	\begin{eqnarray} \label{eq:alpha_vs_r}
	\alpha < (r+\mu) \frac{1}{d_M - \delta}
	\end{eqnarray}}% 
assuming $d_M - \delta > 0$.

The quantity $d_M - \delta$ represents the number of neighbors that are common
to $x_M$ and $x_m$.
Since $\mu \leq r$ it is then sufficient that $d_M - \delta \geq 2$ in order to 
guarantee that $\alpha < r$. Even more, $\alpha$ may become significantly 
smaller than $r$ for large values of $d_M - \delta$. 
Consider for example the case of \emph{complete} graphs.
In this case, $d_M=n-1$, $\delta=0$ and $\mu = r- \alpha$. Hence, 
{\setlength\arraycolsep{1pt}
\begin{eqnarray}
\alpha < \frac{2 r}{n}
\end{eqnarray}}% 
Since $n \geq 2$ we always have $\alpha < r$. Moreover, recalling 
that $r = \max\{\varepsilon,\varepsilon \chi_0 \} + \left( \frac{\eps}{2} + 3 d_{max} \right) |w|_\infty$,
one sees that in the noiseless case $\alpha$ actually 
decreases with $n$ whenever the initial conditions 
do not depend on the network size, and remains bounded irrespective of $w$
with a maximum noise amplification factor equal to $6$.

The considerations made above apply in general since (\ref{eq:alpha_vs_r})
does not depend on the network topology. In fact, (\ref{eq:alpha_vs_r}) suggests
that working with (\ref{eq:local_average}) can be advantageous compared with (\ref{eq:n2n})
whenever the network connectivity is sufficiently large.
We will further substantiate this analysis in Section VII-B through numerical simulations.
 
 \begin{figure*}[!]
 	\centering 
 	\subfigure[State]{\label{fig:smnstate}
 		\includegraphics[width=2.3in]{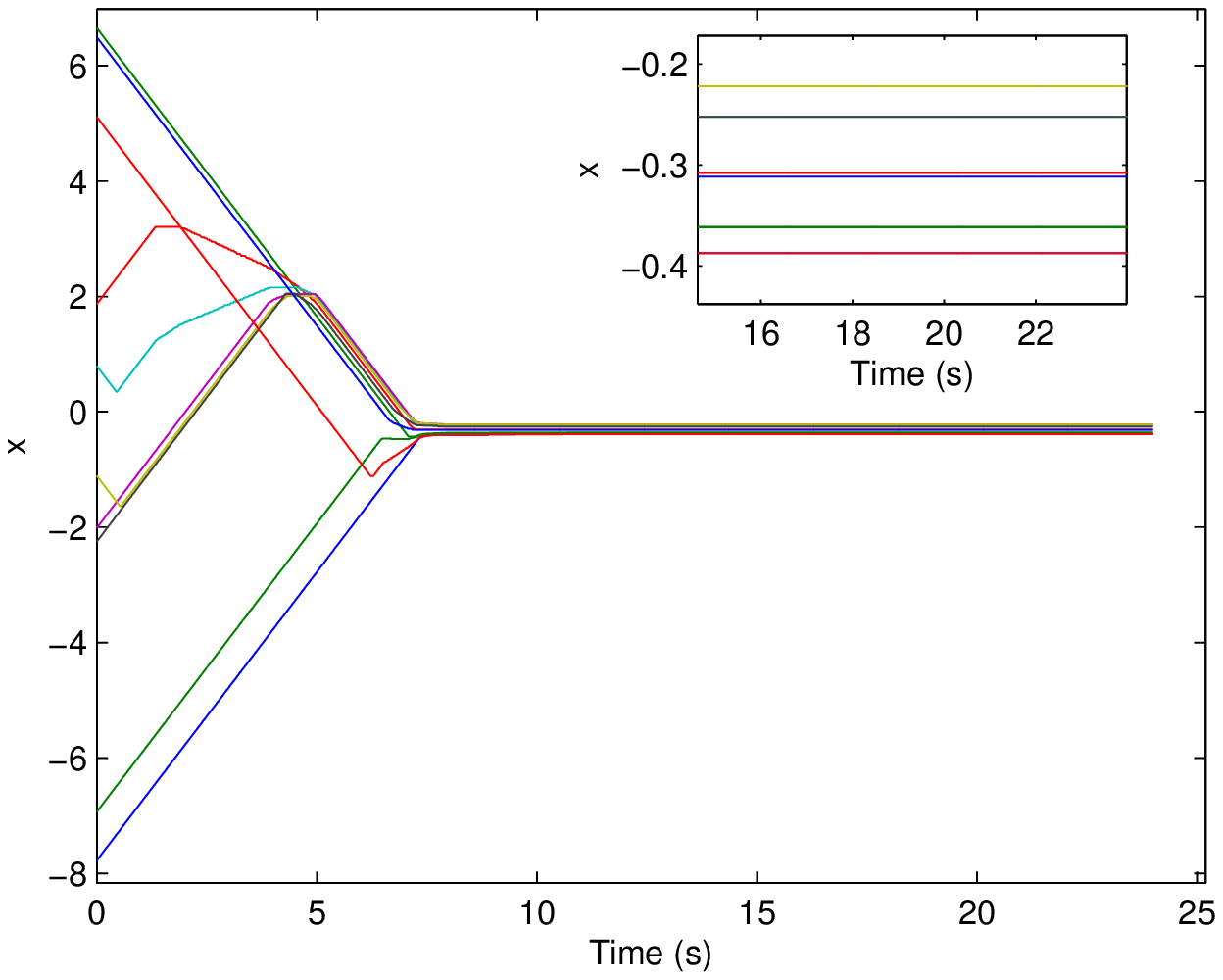}}
 	\subfigure[Absolute value of the noiseless averages]{\label{fig:smnave}
 		\includegraphics[width=2.3in]{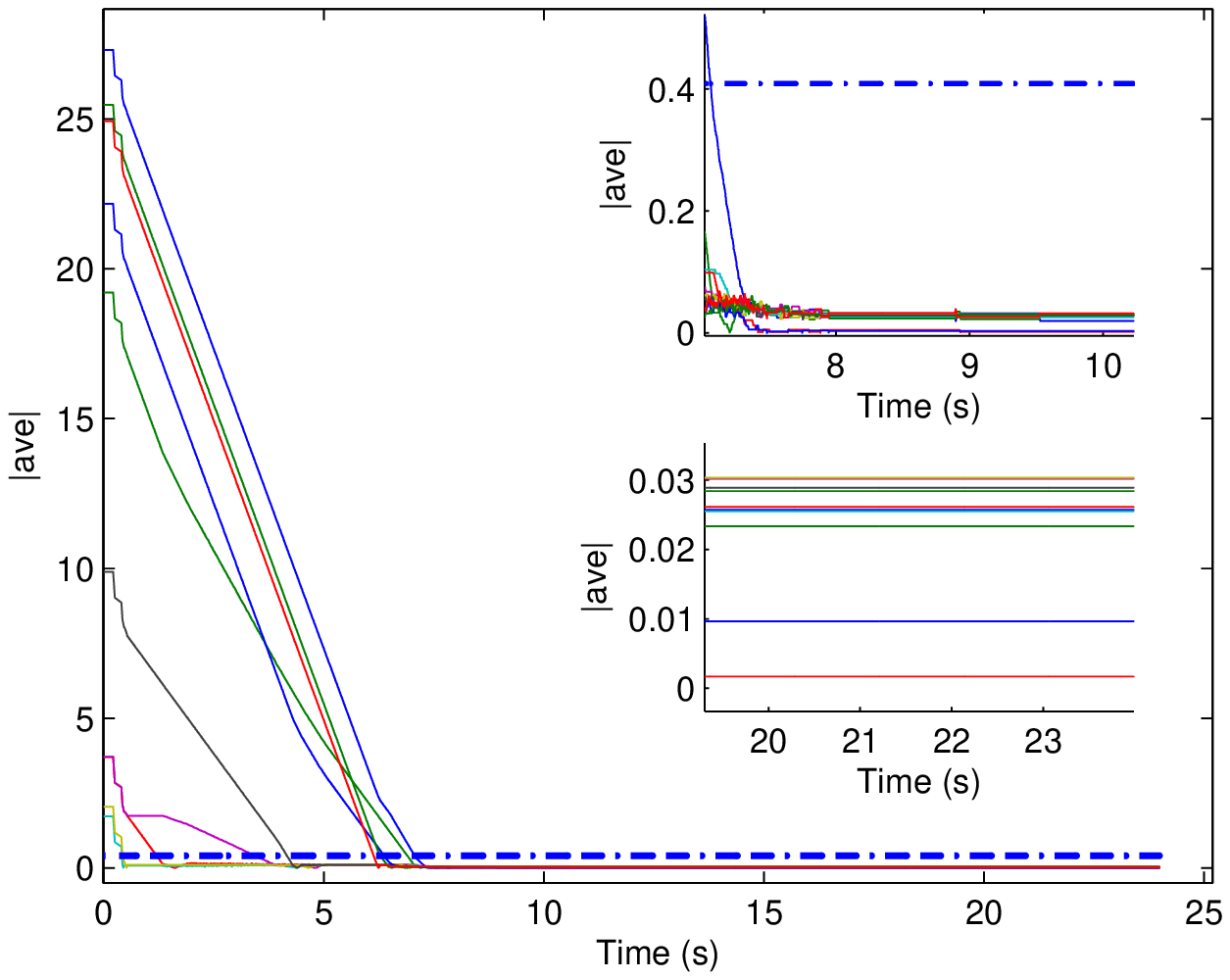}}
 	\subfigure[Local controls]{\label{fig:smnu}
 		\includegraphics[width=2.3in]{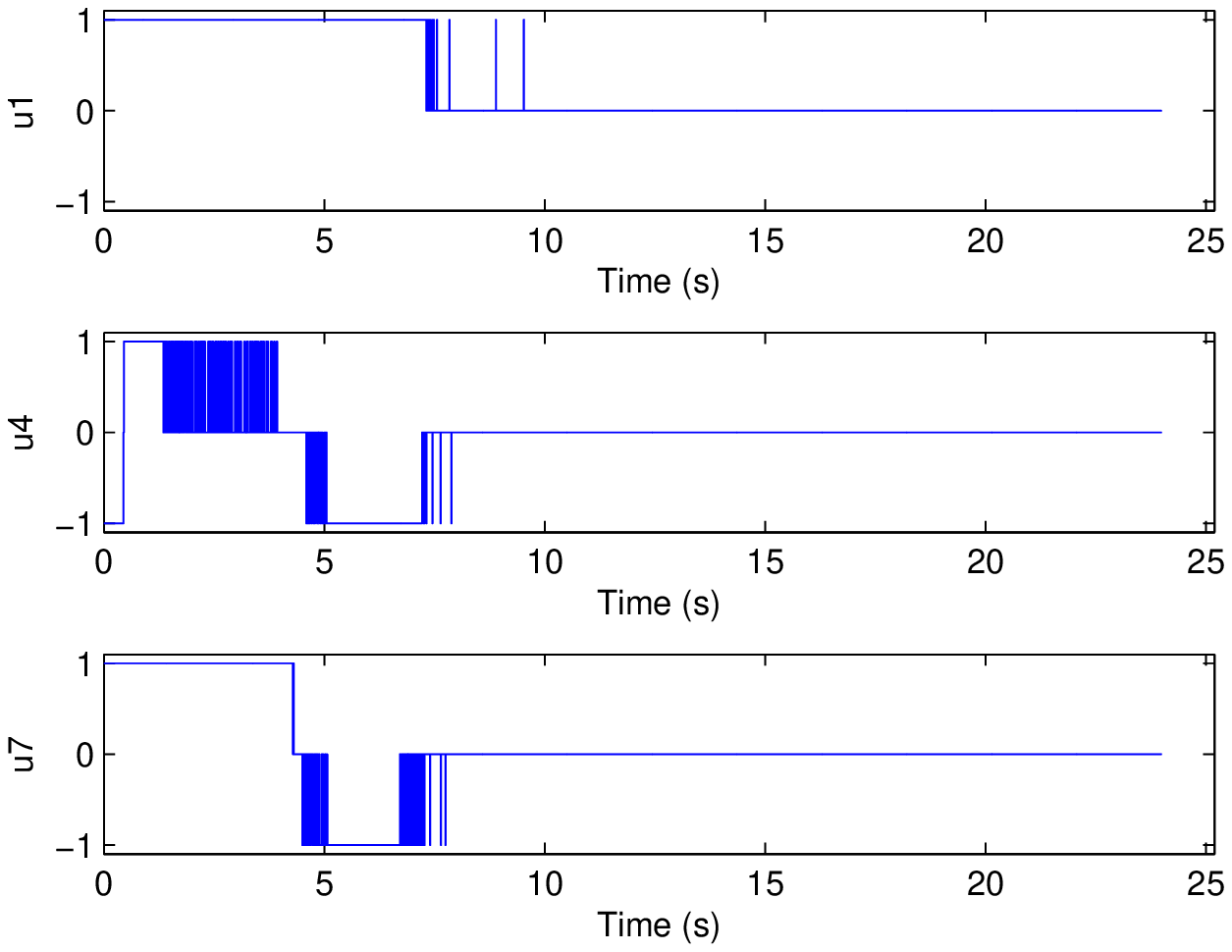}}
 	\caption{Network behavior for $|w|_\infty=0.01$. Since condition $\eps>2d_{max}|w|_\infty$
	is satisfied, then the network state eventually converges
	to a point belonging to the set $\mathcal D$ in (\ref{eq:setD}) (Theorem \ref{thm:smallnoise}). 
	Moreover, the state remains confined in the initial envelope 
	(Theorem \ref{thm:statebounded}).} \label{fig:smn}
 \end{figure*}      

 \begin{figure*}[!]
 	\centering 
 	\subfigure[State]{\label{fig:lgnstate2}
 		\includegraphics[width=2.3in]{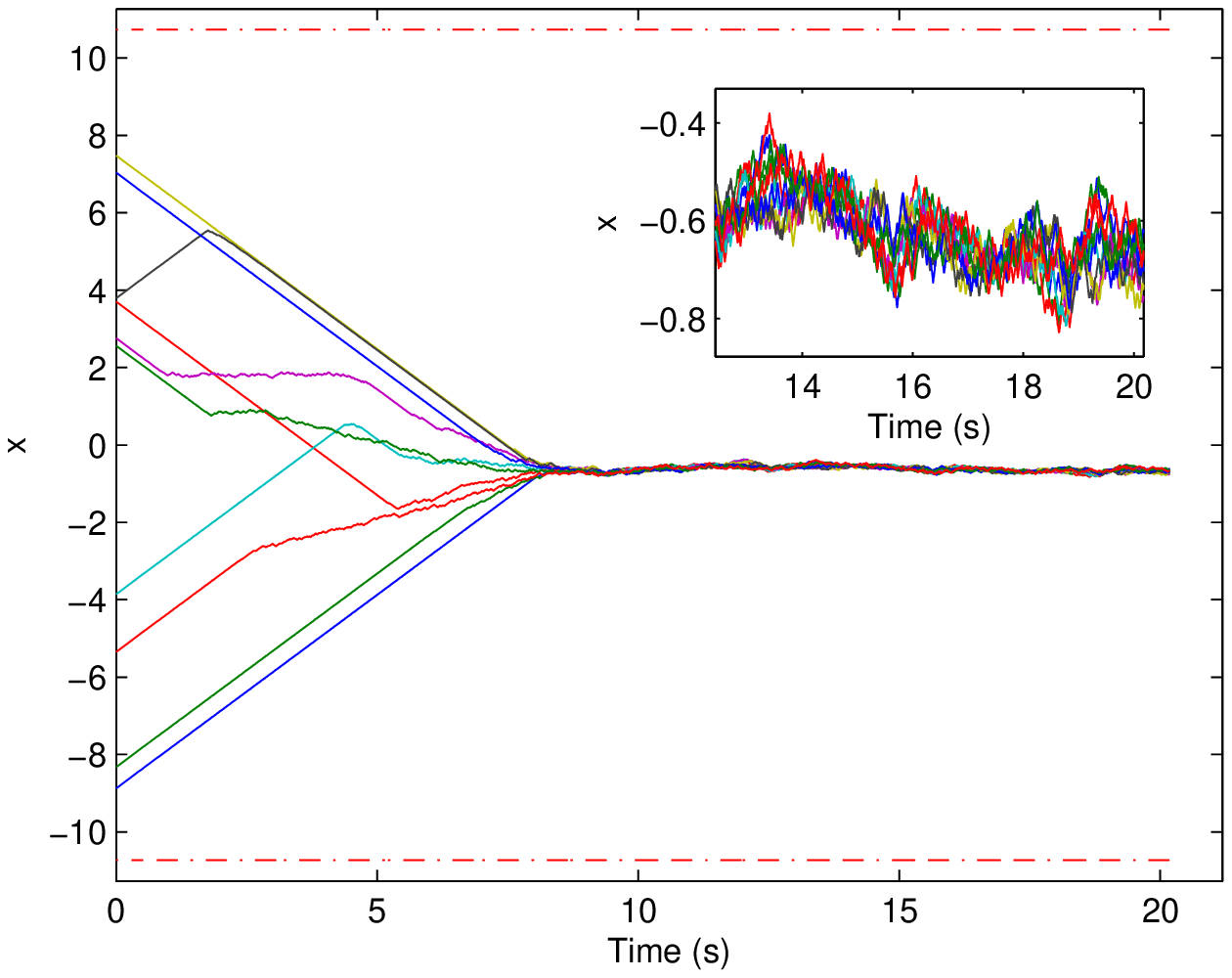}}
 	\subfigure[Absolute value of the noiseless averages]{\label{fig:lgnave2}
 		\includegraphics[width=2.3in]{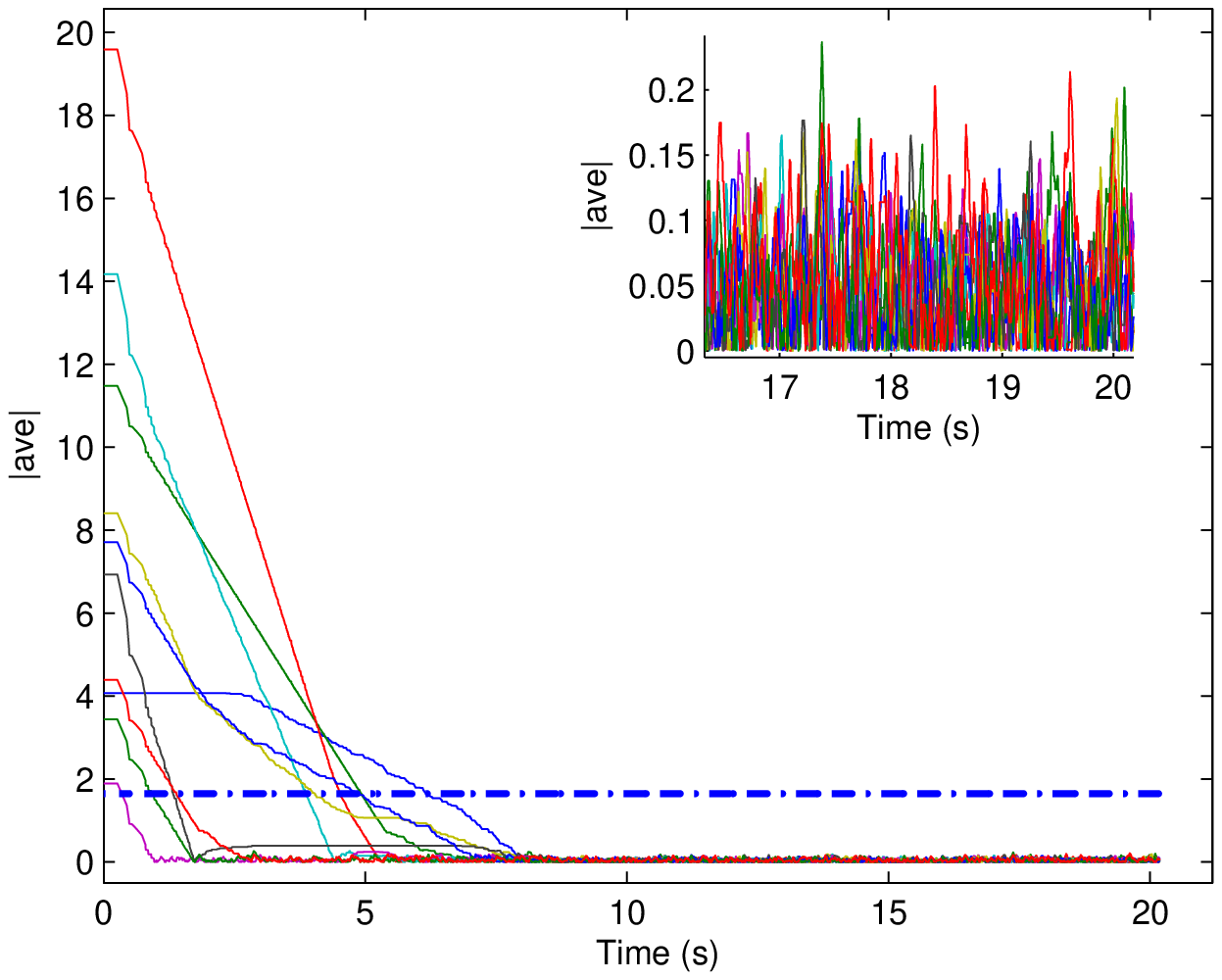}}
 	\subfigure[Local controls]{\label{fig:lgnu2}
 		\includegraphics[width=2.3in]{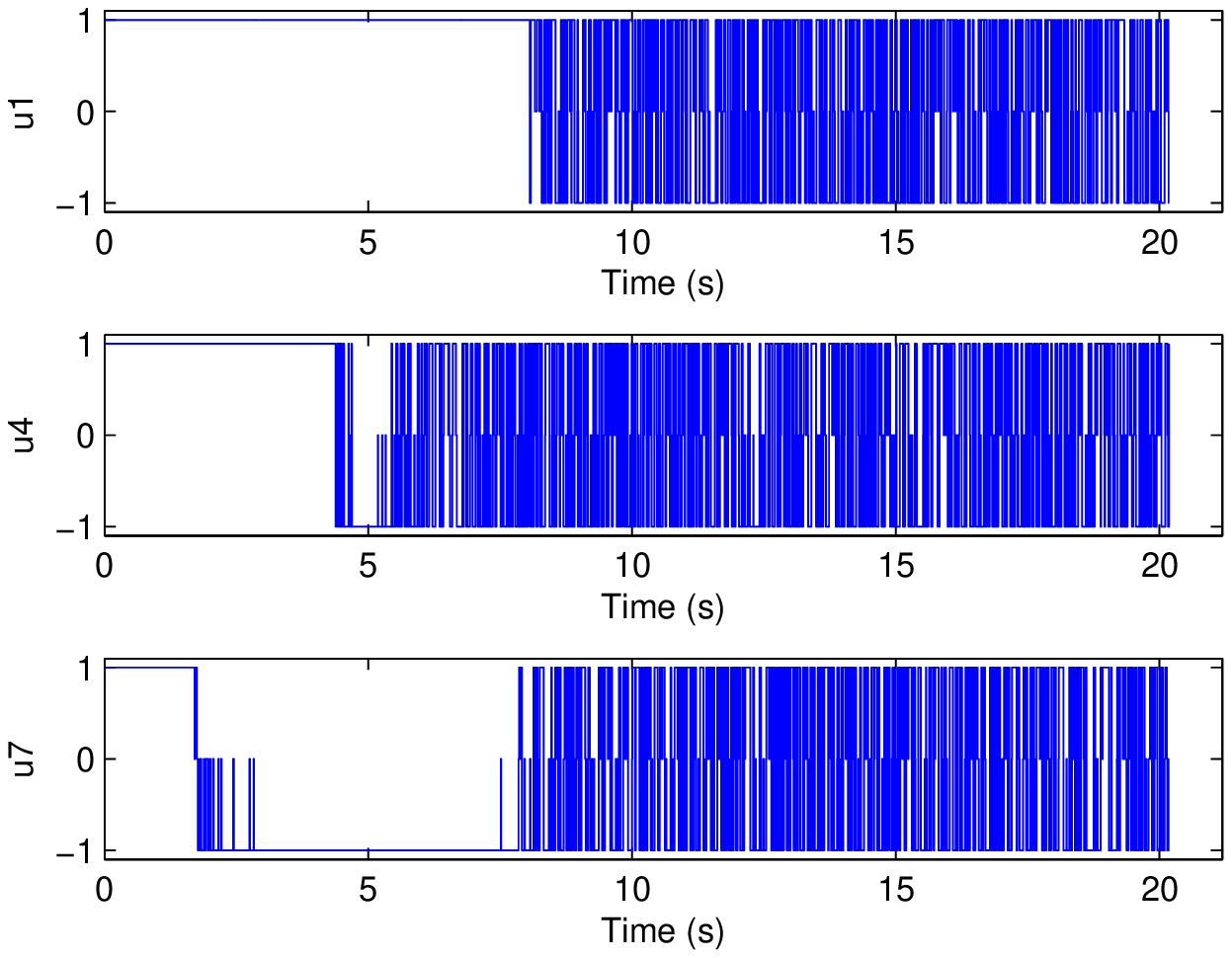}}
 	\caption{Network behavior for $|w|_\infty=0.2$. 
	Condition $\eps>2d_{max}|w|_\infty$ is not satisfied 
	and the state continues to fluctuate inside
	$\mathcal D$ (Theorem \ref{thm:generalnoise}). 
	} \label{fig:lgn2}
 \end{figure*}
 
  \begin{figure*}[!] 
 	\centering 
 	\subfigure[State]{\label{fig:lgnstate}
 		\includegraphics[width=2.3in]{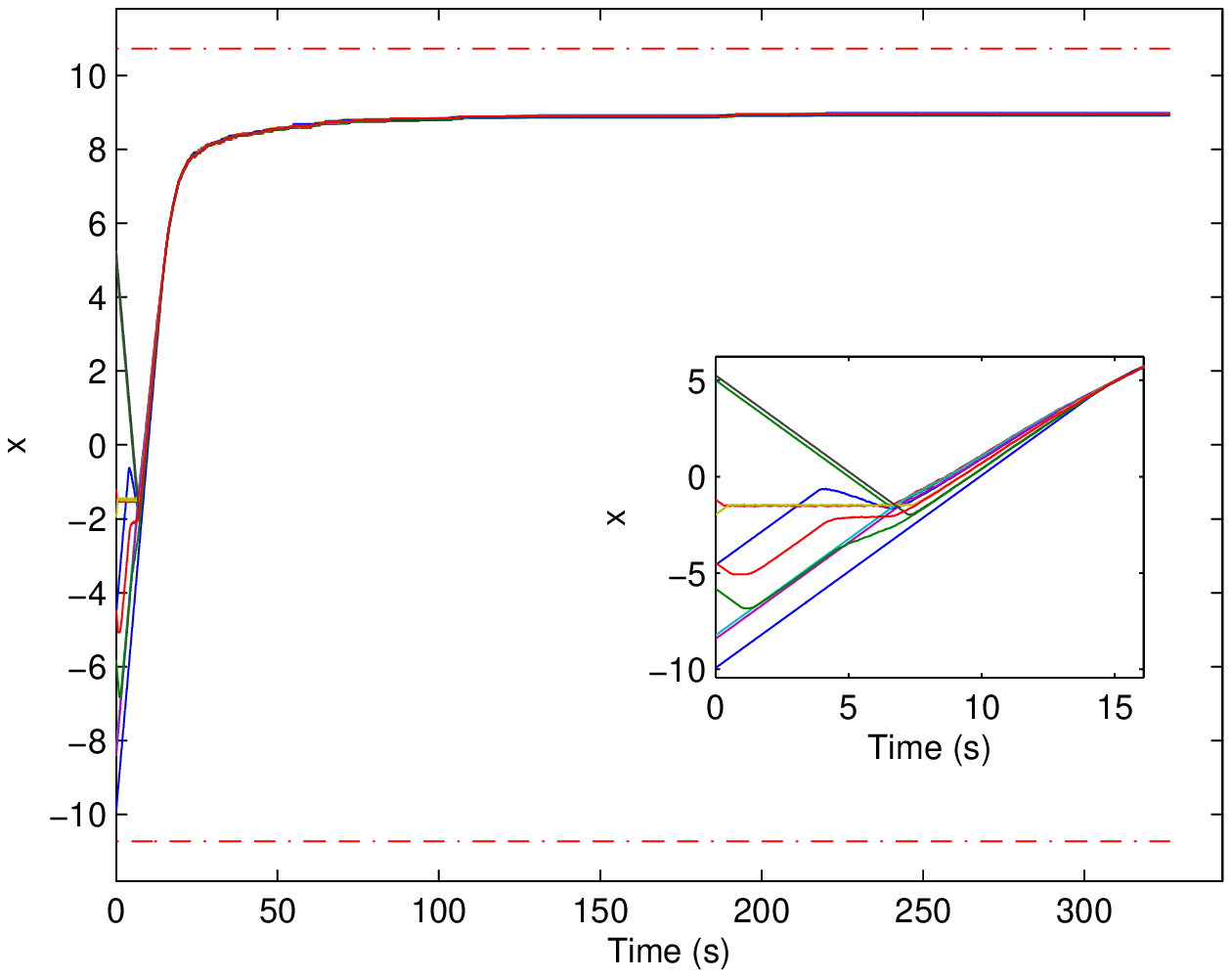}}
 	\subfigure[Absolute value of the noiseless averages]{\label{fig:lgnave}
 		\includegraphics[width=2.3in]{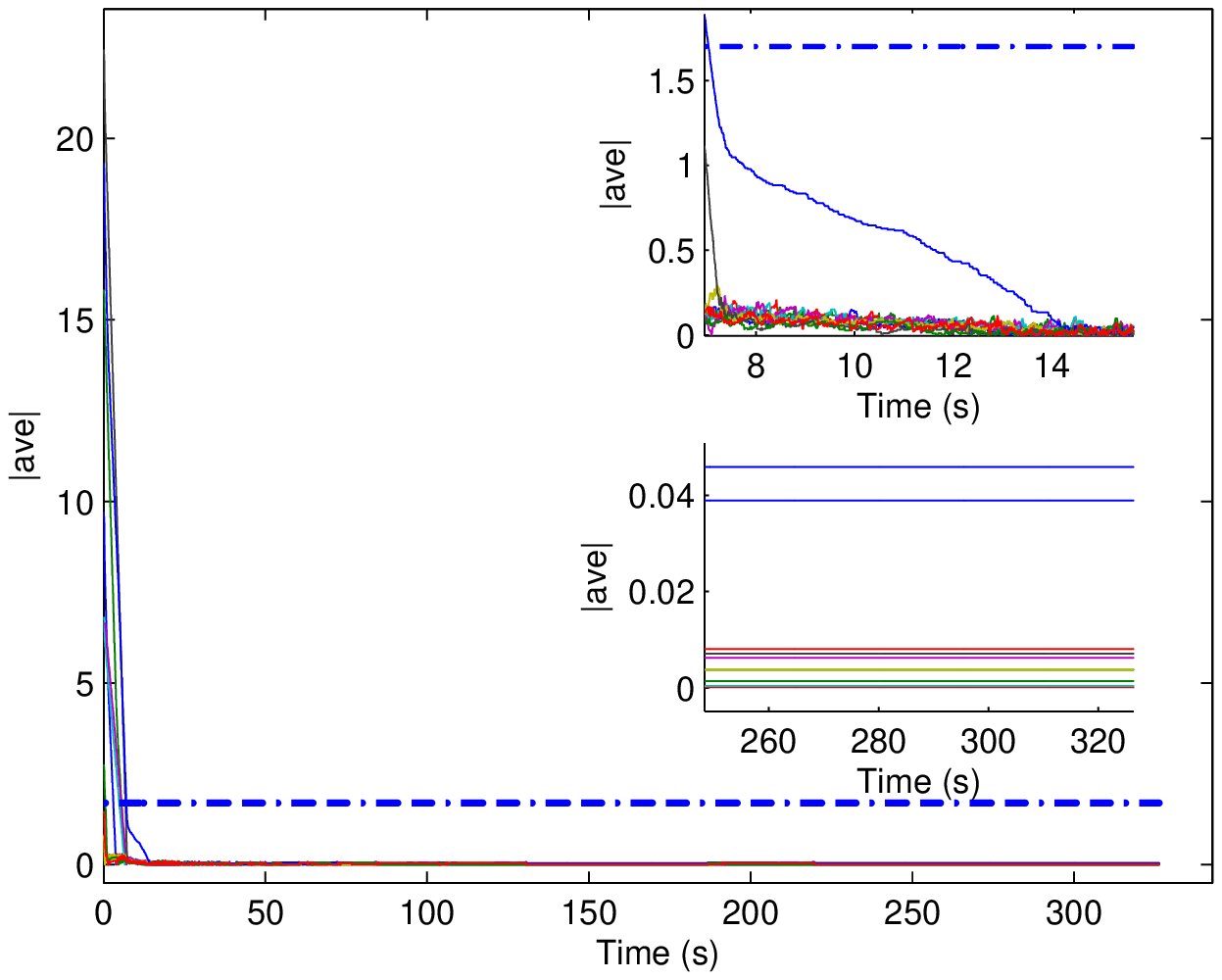}}
 	\subfigure[Local controls]{\label{fig:lgnu}
 		\includegraphics[width=2.3in]{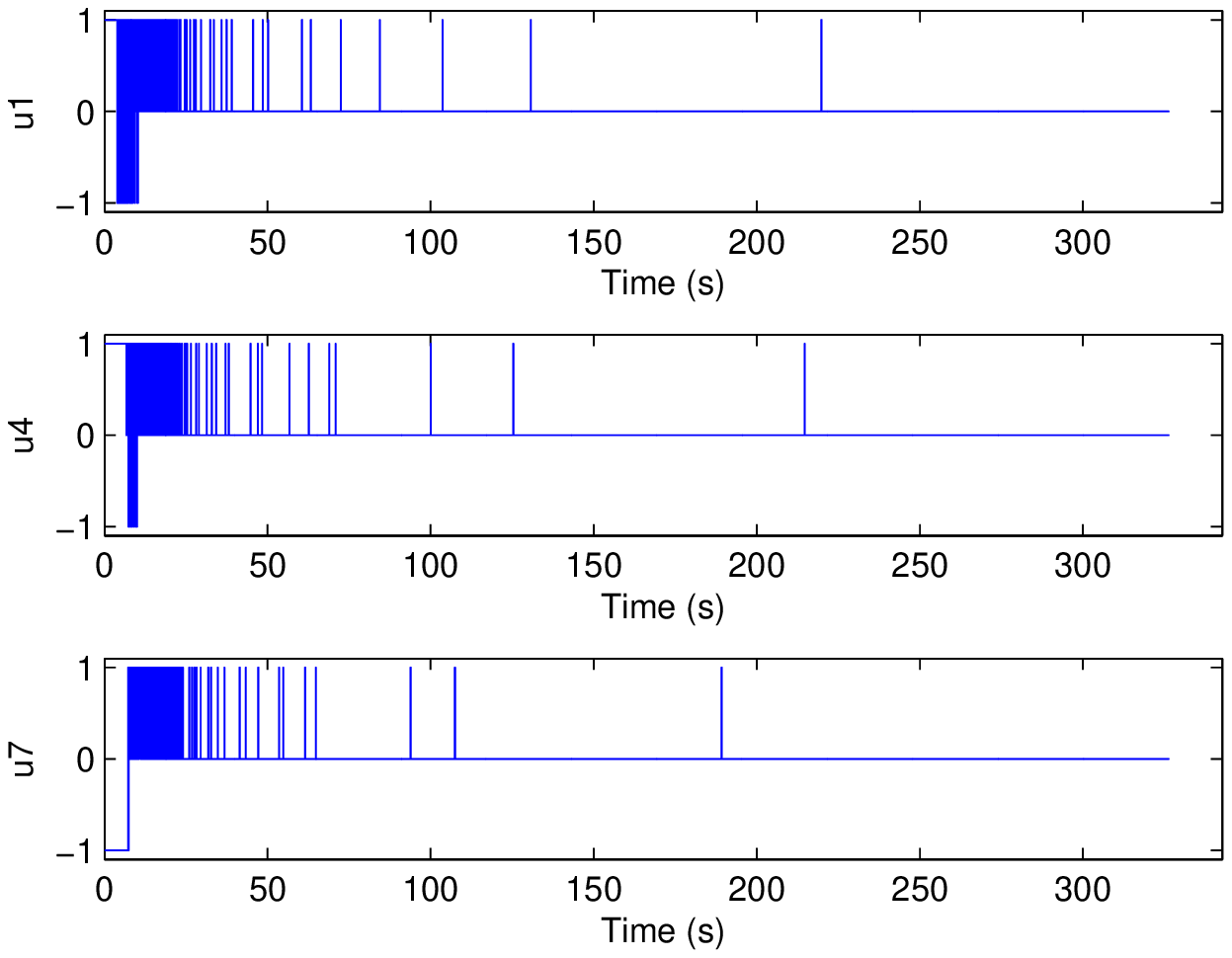}}
 	\caption{Network behavior for $|w|_\infty=0.2$ with sign-preserving noise. 
	The state initially drifts but the drifting eventually stops thanks to the adaptive threshold 
	mechanism. Condition $\eps>2d_{max}|w|_\infty$ is not satisfied and  
	the state does not remain within the initial envelope 
	(Theorem \ref{thm:statebounded}).} \label{fig:lgn1}
 \end{figure*}

\section{Numerical Examples} \label{sec:examp}

In this section, we illustrate the proposed consensus scheme 
through a number of numerical examples.

\subsection{Small graph} \label{Exm:smallgraph} 

This example is used to illustrate the main results of this paper 
in an easy-to-follow manner. 
We consider a simple cycle graph with $10$ nodes, which implies $d_{max}=2$. 
Moreover, we let $\eps=0.05$. The initial value of each network node
is taken as a random number within $[-10,10]$. 

\emph{Low-magnitude noise.}
To begin with, we assume that the noise  are generated randomly 
within $[-0.01,0.01]$, which implies $\eps>2d_{max}|w|_\infty$. 
The simulation results are reported in Figure \ref{fig:smn}, which
shows trajectory of the states $x_i$, absolute values of local averages $|\ave_i|$,  
and local controls for nodes $1$, $4$ and $7$. 
One sees that the conditions of Theorem \ref{thm:smallnoise} 
are verified in the sense that the network state eventually converges 
and the local controls become zero, which occurs after $\approx10s$. 
In Figure \ref{fig:smnave}, the blue dot-dash line represents the bound on
$r$ dictated by Theorem \ref{thm:smallnoise}. 
In this example, $r=0.41$. 
Moreover, by Theorem \ref{thm:statebounded} the state evolution 
remains confined in the initial envelope since $\chi_0 \approx 7.8 > \gamma \approx 0.5366$.  

\emph{General case: Zero mean noise.}
We next assume that the noise for node $i$ is given by
\begin{eqnarray}
w_i(t)=v_i(t)+0.04\times\sin(2it+i\pi/(3n)) 
\end{eqnarray}
where $v_i$ is generated randomly within $[-0.16, 0.16]$ and {\color{black}$n=10$}. This implies 
$|w|_\infty=0.2$ so that $\eps < 2d_{max}|w|_\infty$. 
 Simulation results are shown in Figure \ref{fig:lgn2}, from which one sees 
that the state enters the set $\mathcal{D}$ around $t\approx6.2s$ and remains there forever, 
while the local controls continue to switch. 
This is in agreement with Theorem \ref{thm:generalnoise}, 
as well as the discussion in Remark \ref{rem:nonstatic}.

\emph{General case: Sign-preserving noise.}
We finally assume that the noise  are generated randomly 
within $[0,0.2]$, which implies again $\eps < 2d_{max}|w|_\infty$. 
Since the Laplacian has an eigenvalue in zero, constant 
or sign-preserving noise  represent a critical situation since they can induce 
drifting phenomena. This phenomenon is shown in Figure \ref{fig:lgn1}.
One sees that the proposed coordination scheme prevents the state from
growing unbounded. In particular, in agreement with Theorem \ref{thm:statebounded}
the state remains within the interval $[-\gamma,\gamma]$ with $\gamma \approx 10.73$
(red dot-dash line in Figure \ref{fig:lgnstate}).
In agreement with Theorem \ref{thm:generalnoise},
the network state enters in a finite time the set $\mathcal D$ and remains there forever. 
Figure \ref{fig:lgnave} shows that the theoretical bound $r\approx1.7$ (blue dot-dash line) is conservative 
as each local average eventually becomes very small.
From Figure \ref{fig:lgnu} one sees that the local controls do not switch as fast as in the beginning. 
This is expected since, as state increases, also the threshold increases. This makes the noisy average
$\ave^w_i$ likely to be confined within $(-\eps_i,\eps_i)$, causing
the control switches to be more and more sporadic.

  \begin{figure*}[!]
  	\centering 
  	\subfigure[$A_{MLA}$]{\label{fig:A_mla}
  		\includegraphics[width=2.3in]{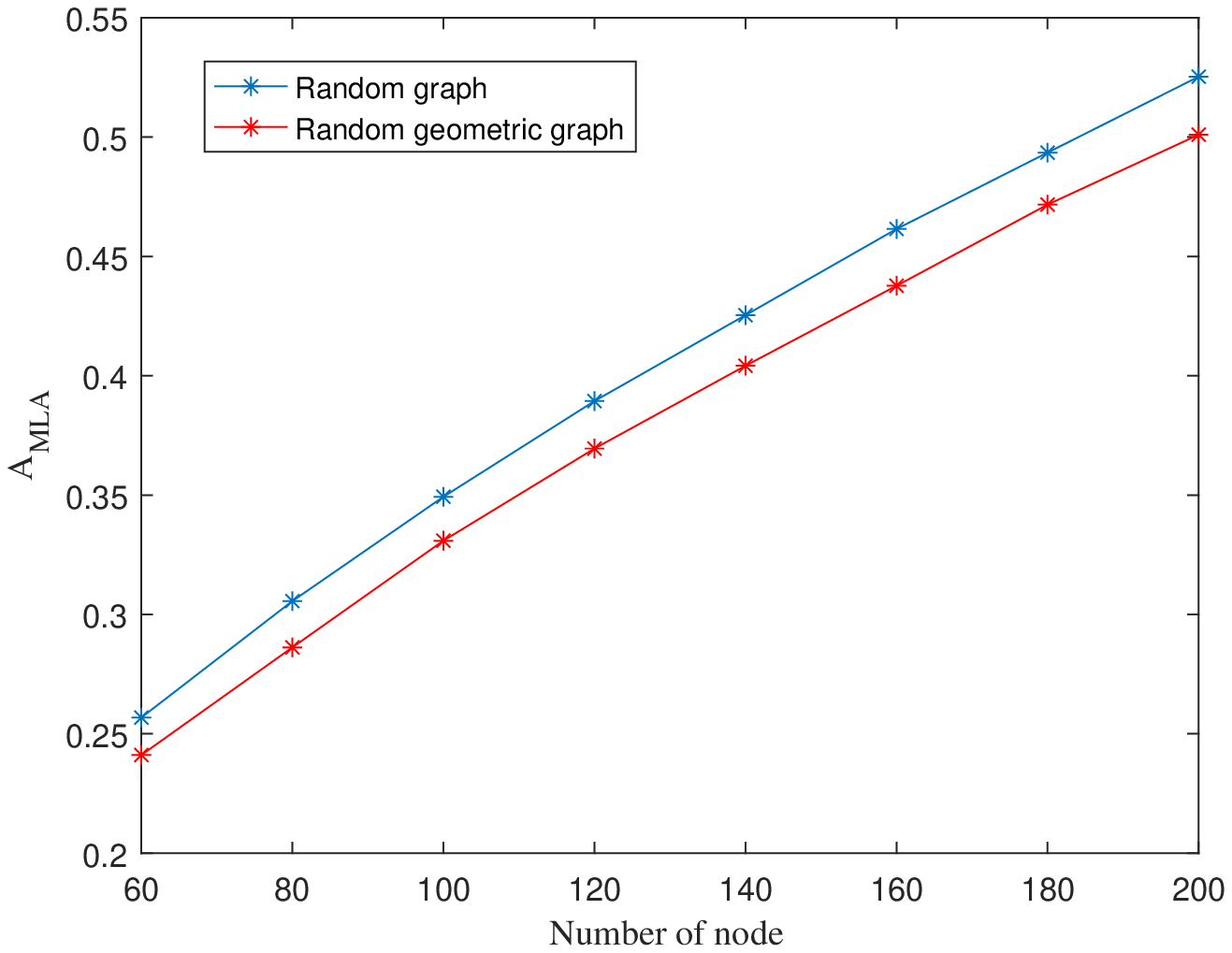}}
  	\subfigure[$A_{MND}$]{\label{fig:A_mnd}
  		\includegraphics[width=2.3in]{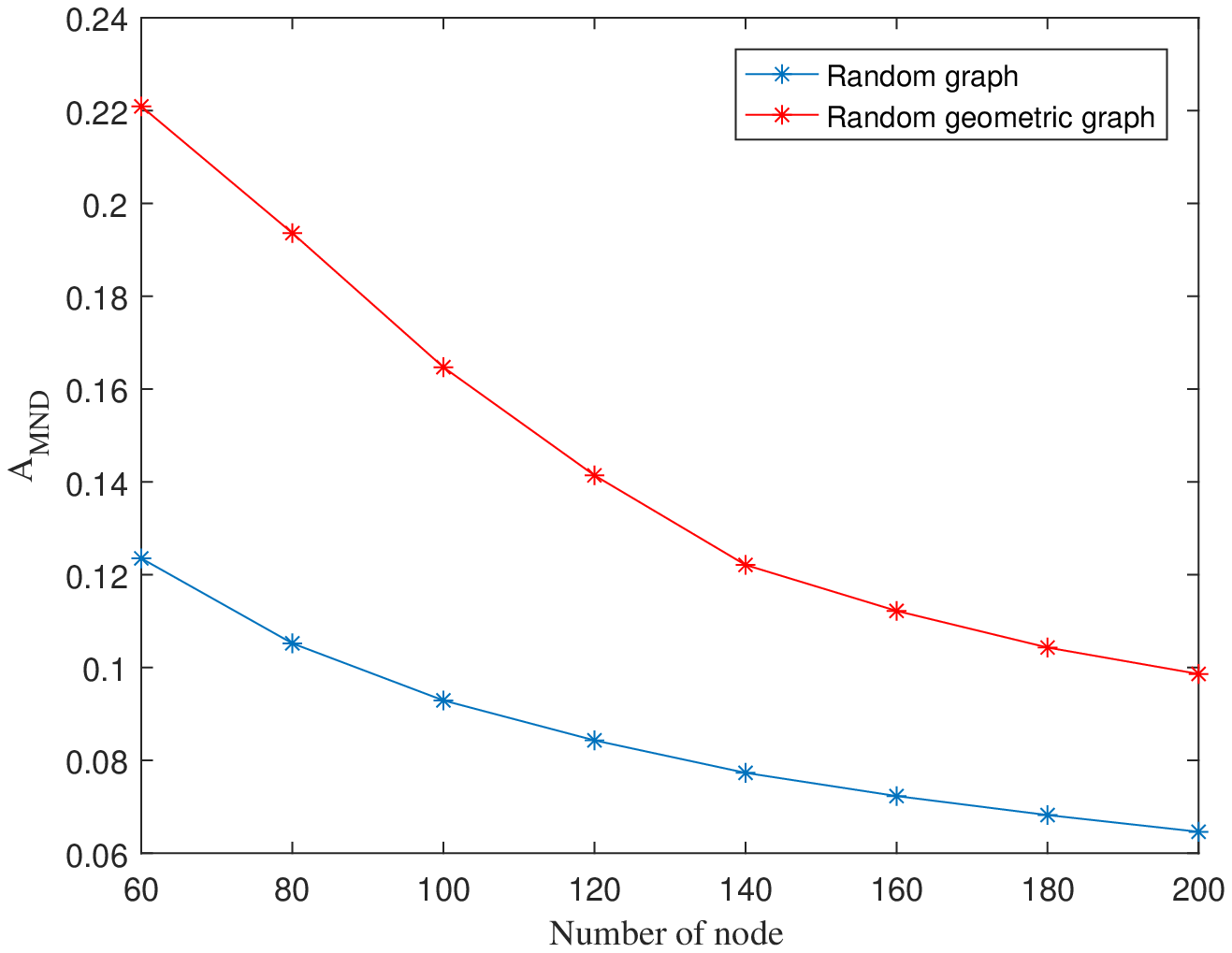}}
  	\subfigure[$A_{MDEC}$]{\label{fig:A_mdec}
  		\includegraphics[width=2.3in]{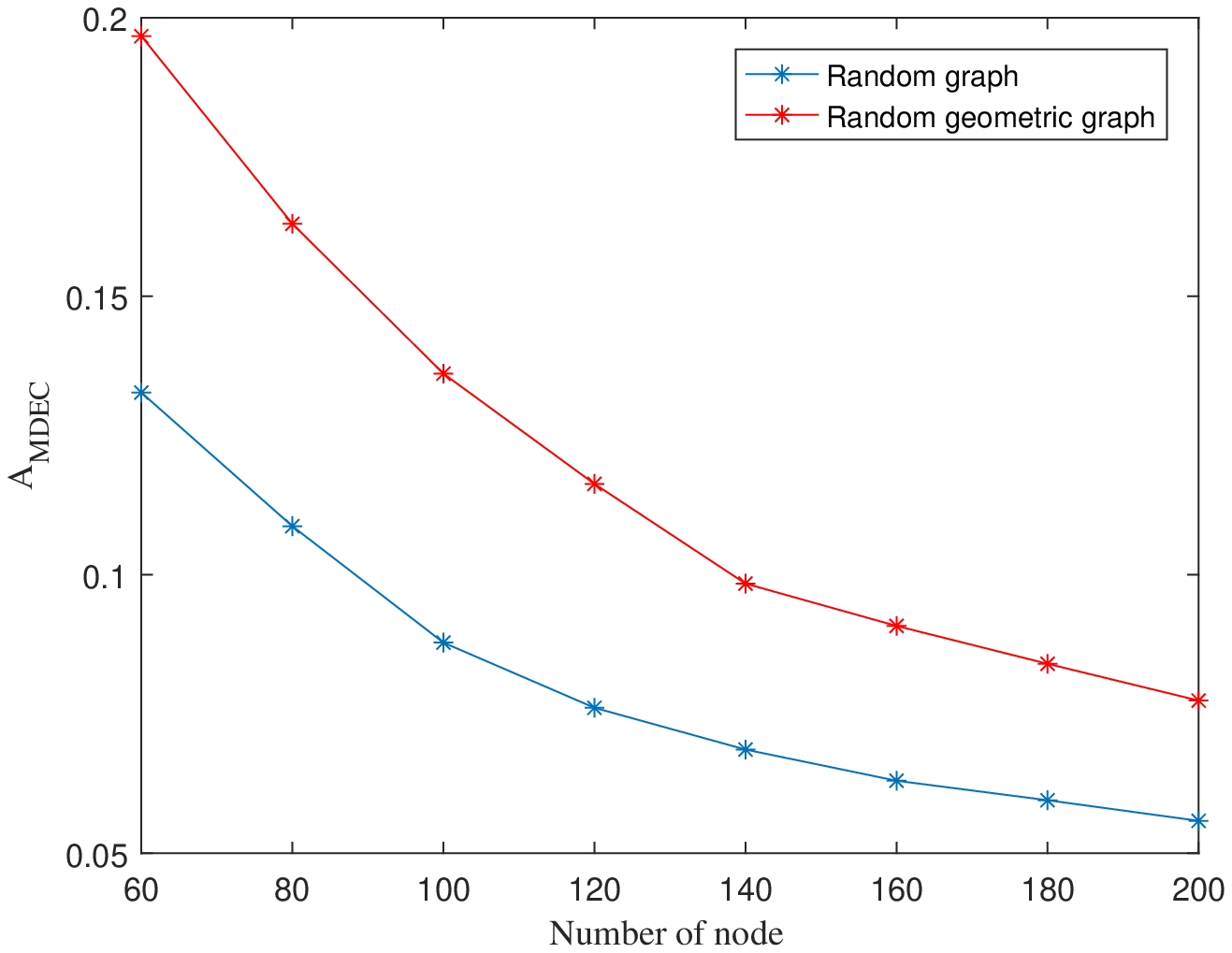}}
  	\caption{Monte Carlo simuation results for the Erd\"os-R\'{e}nyi (ER) and 
	random geometric (RG) graphs}\label{fig:Montecarlo}
  \end{figure*}

\subsection{Erd\"os-R\'{e}nyi and random geometric graphs}

In this section, we illustrate the proposed scheme for graphs of a larger size
and exemplify some of the considerations made in Section VI-B,
focusing on two well-known graphs: Erd\"os-R\'{e}nyi (ER) and random geometric (RG) graphs \cite{Penrose}.
The former is obtained from the $n$-dimensional complete
graph by retaining each edge with probability $p$ (independently).
The latter is obtained by considering a random uniform deployment of $n$ points 
in a $2$-dimensional Euclidian space. Denoting by 
$s_i$ the position of node $i$, a link between nodes $i$ and $k$ 
exists if and only if $| s_i - s_k | \leq R$ where $R$ denotes
the communication range, which is assumed identical for every node.

For both the graphs we consider Monte Carlo simulations.
Specifically, we consider $N_{trials}=1000$ trials.
For each trial, we generate an ER (RG) graph 
of $100$ nodes. Graphs which are not connected are not taken into account.  
For the ER graph we consider a link probability $p=0.08$, while for the RG graph
we consider a random deployment over a region of $1km \times 1km$
with nodes communication range $R=160m$.  
For each trial, the nodes initial values are taken randomly within $[-2,2]$,
and the noise is taken as a random number within $[-0.2,0.2]$. 
The sensitivity parameter is $\eps=0.1$ for all the trials.

Let $\{t_s\}_{s \in \mathbb N_0}$ be the sequence of time instants
at which one of the nodes samples, \emph{i.e.} $t_s=t^i_k$ for some
$i \in I$ and $k \in \mathbb N_0$. Given a simulation horizon $H$, 
this sequence will range from $t_0$ up to $t_S$ where $S$ is the largest 
integer such that $t_S \leq H$. The \emph{asymptotic} behavior 
of the nodes is defined as the behavior of the nodes over the time interval 
$[t_{S-W+1},t_{S-W+2},\ldots,t_{S}]$, where $W$ is a positive integer that 
is selected so as to satisfy $W \gg1$ and $W \ll S$. 
 The reason for this choice is twofold: (i) since the network nodes need not converge, it makes little 
 sense to consider only the value of the nodes at the final step $t_S$.
 In this respect, $W \ll S$ makes it possible to evaluate the network behavior 
 for a sufficiently large number of samples; (ii) we aim at 
 evaluating the network limiting behavior, \emph{i.e.} after the transient 
 has vanished. Hence, $W \gg 1$ guarantees that initial samples are not 
 taken into account. In the simulations, for each trial, we consider, $H=10^5$
 and $W=1000$.
 We consider three performance indices:
 \begin{enumerate}
 	\item \emph{Asymptotic maximum local average.} 
 	This index is given by
 	\begin{eqnarray}
 	&& A_{MLA} := \nonumber \\ 
	&& \quad \frac{1}{N_{trials}} \sum_{k=1}^{N_{trials}} 
 	\left(  \frac{1}{W} \sum_{s=S-W+1}^{S}  \max_{i \in I} |\ave_i(t_s)| \right)
 	\nonumber
 	\end{eqnarray}
 	Basically, for each of the trials, we compute the average of the largest value 
 	of the local averages over the time interval $[t_{S-W+1},t_{S-W+2},\ldots,t_{S}]$.
 	Then, these values are averaged over the number of trials.
 	\item \emph{Asymptotic maximum node-to-node distance.} 
 	This index is given by 
 	{\setlength\arraycolsep{1pt}
 		\begin{eqnarray}
 		&& A_{MND} := \nonumber \\
 		&& \,\,\,  \frac{1}{N_{trials}} \sum_{k=1}^{N_{trials}} 
 		\left(  \frac{1}{W} \sum_{s=S-W+1}^{S}  \max_{i,j \in I} |x_i(t_s)-x_j(t_s)| \right)
 		\nonumber
 		\end{eqnarray}}%
 	Here, for each trial, 
 	we compute the average of the largest value of the node-to-node distances 
 	over the interval $[t_{S-W+1},t_{S-W+2},\ldots,t_{S}]$.
 	As before, these values are then averaged over the number of trials.
	 	\item \emph{Asymptotic maximum distance from the expected convergence point.} 
 	This index is given by
 	{\setlength\arraycolsep{1pt}
 		\begin{eqnarray}
 		&& A_{MDEC} := \nonumber \\
 		&& \quad \frac{1}{N_{trials}} \sum_{k=1}^{N_{trials}} 
 		\left(  \frac{1}{W} \sum_{s=S-W+1}^{S}  \max_{i \in I} |x_i(t_s)-x_*| \right)
 		\nonumber
 		\end{eqnarray}}%
 	where
 	\begin{eqnarray}
 	x_* := \frac{\max_{i \in I} x(0) + \min_{i \in I} x(0)}{2}
 	\end{eqnarray}
This performance index is similar to $A_{MND}$, with the exception
 that the nodes values are compared to the midpoint $x_*$ of the maximum 
 and minimum initial values of the nodes. This is because our algorithm 
 can be viewed as an approximation of the pure $\sign(\ave_i)$-consensus,
 which is known to converge to $x_*$ \cite{cortes2006finite}. 
 \end{enumerate} 
 
The results are reported in Figure \ref{fig:Montecarlo}. Figure \ref{fig:A_mla} confirms 
the bound obtained in Theorem \ref{thm:generalnoise}, showing that the local averages
scale nicely with $d_{max}$ (\emph{cf.} (\ref{eq:r})). More interesting is the result in 
Figure \ref{fig:A_mnd} which shows that the node-to-node error decreases 
as the number of nodes increases.  
This can be explained by observing that for both the graphs the expected number 
of common neighbors increases with $n$, which causes $\alpha$ in (\ref{eq:alpha_vs_r})
to decrease in agreement with the comments made in Section VI-B. 
In particular, for the ER graph the expected number 
of common neighbors between two network nodes is given by $(n-2)p^2$, while 
for the RG graph the probability that two nodes are connected is given 
by $\bar p = \pi R^2/ |A|=0.08$ where $|A|$ is the area of the deployment region,
and the expected number of common neighbors between two connected nodes 
is approximately $0.58 n \bar p$ \cite{Chan}. This can explain why $A_{MND}$ is smaller for the ER graph.
Figure \ref{fig:A_mdec} finally shows that the distance from the expected convergence point 
is indeed small and decreases with $n$. The latter property can be explained 
by noting that large values of $n$ decrease the effect of $\eps$ (\emph{cf.} Section VI-B), which causes 
the quantized sign function to better approximate the pure $\sign(\ave_i)$ function.

 \begin{figure}
	\centering 
	\includegraphics[width=0.5 \textwidth]{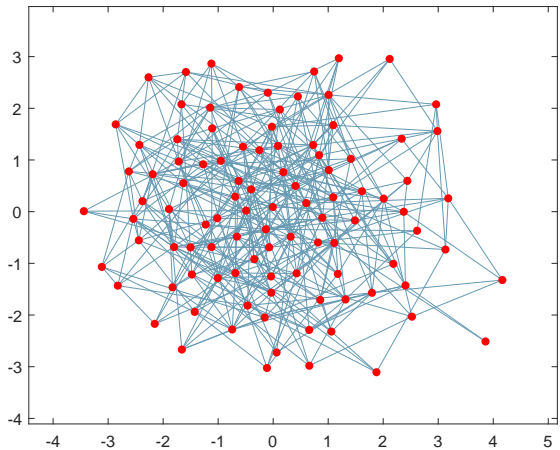}
	\caption{Network topology for one of the trials for the ER graph}
	\label{fig:Rdgraph}
\end{figure}
  
We report in Figures \ref{fig:Rdgraph} and \ref{fig:Rdgraph_Sim} the results of 
one of the trials for the ER graph. In this trial, we obtain $d_{max}=14$
which leads to $r=8.8067$ and $\gamma=37.2667$. The large theoretical 
bounds are due to the large value of $d_{max}$. 
In practice, as show in Figure \ref{fig:Rdgraph_Sim}, the regulation performance is very high.
In fact, the absolute value of the noiseless averages is eventually 
upper bounded by $0.5$, which is much smaller than the theoretical bound
given by $r$. We omit the simulation results of one trial for the RG graph 
since the figures are similar to the ones for the ER graph.

 \begin{figure}
 	\centering 
 	\subfigure[State]{\label{fig:state_rdgraph}
 		\includegraphics[width=0.5 \textwidth]{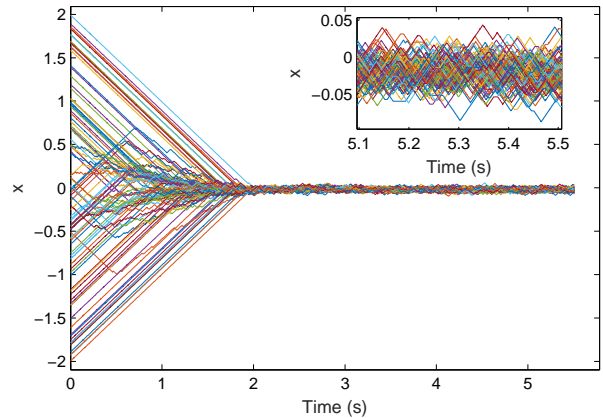}}
 	\subfigure[Absolute value of the noiseless averages]{\label{fig:ave_rdgraph}
 		\includegraphics[width=0.5 \textwidth]{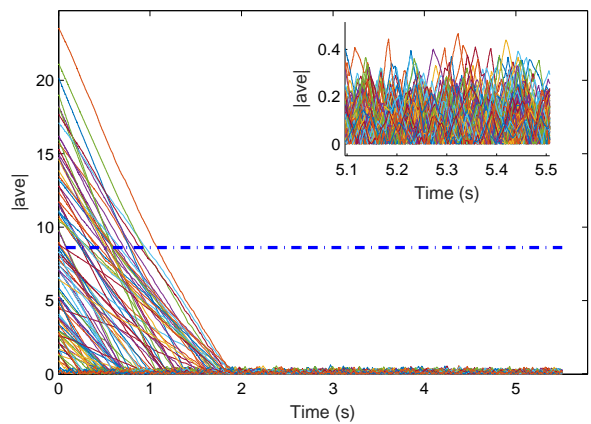}}
 	\caption{Network behavior for one of the trials for the ER graph.}
	\label{fig:Rdgraph_Sim}
 \end{figure}

\section{Conclusions}

\label{sec:conclusion}
In this paper, we proposed a novel self-triggered network coordination scheme 
that can handle unknown-but-bounded noise  affecting the network 
communication. The proposed coordination scheme employs a dynamic, 
state-dependent, triggering policy and ternary controllers. 
It has been shown that the scheme can achieve finite-time practical consensus in both noiseless and 
noisy cases. In the latter situation, the node disagreement value scales nicely with the magnitude 
of the noise. An interesting feature of the proposed scheme is that 
the implementation does not require any global information 
about the network parameters and/or the operating environment.
Moreover, the communication between nodes 
occurs only at discrete time instants, and nodes can sample independently and in an aperiodic manner.
The last feature renders the proposed scheme applicable when coordination is through 
packet-based communication networks. 

An interesting outcome of this work is that the proposed scheme 
can guarantee a small node-to-node error without requiring to choose the consensus threshold 
too small. In turn, this can be beneficial for moderating the noise sensitivity as well as the number of 
communications. Investigating this point in more details certainly represents an interesting 
venue for future research.

\appendix

\section*{Communication Delays}
 
In this section, we briefly discuss how transmission delays can be taken into account.
Some of the derivations follow closely the delay-free analysis of Section III-B so that 
we will discuss in detail only the points where substantial differences appear.

For each $i \in I$, let $\{t_k^i\}_{k \in \mathbb N_0}$ with $t^i_0=0$ be
the sequence of time instants at which node $i$ starts collecting
data from its neighbors. Given a neighbor $j \in \mathcal N_i$, node $i$
will receive information from $j$ at a certain time $s_k^{ij} := t_k^i + \tau_k^{ij}$,
where $\tau_k^{ij}$ represents the total delay in the communication between $i$ and $j$.
In general, $\tau_k^{ij}$ can be time-varying (dependence on $k$) as well as link-dependent (dependence on $i$ and $j$).
At $s_k^{ij}$, the information received by node $i$ is given by
$z_j(v_k^{ij})$ for some $v_k^{ij} \in [t_k^i,s_k^{ij}]$, which represents
the time at which $j$ transmits its value. 
At time
\begin{eqnarray} \label{eq:local_total_delay}
s_k^{i} := \max_{j\in\neigh{i}} s_k^{ij} = t^i_k + \max_{j\in\neigh{i}} \tau_k^{ij}
\end{eqnarray}
node $i$ will then have all the information needed to update its control
action. Accordingly, $\{s_k^i\}_{k \in \mathbb N_0}$
will define the sequence of control updates.

The control action is given by
\begin{eqnarray} \label{eq:controls_delay}
u_i(t) =
\left\{
\begin{array}{ll}
0 & \quad t \in [0,s^i_0[ \\ \\
\displaystyle \ave^{w,\tau}_i(s^i_k) & \quad
t \in [s^i_k,s^i_{k+1}[  
\end{array} \right.
\end{eqnarray}
where
\begin{eqnarray} \label{eq:ave_noisy_delay}
\ave^{w,\tau}_i(s^i_k) := \sum_{j\in\neigh{i}}(z_j(v_k^{ij})-x_i(s^i_k))
\end{eqnarray}
The rationale is the following.
Before time $s_0^i$, node $i$ has no information from the whole neighboring set 
so that its control action is set to zero. On the other hand, $\ave^{w,\tau}_i$
is nothing but the natural generalization of the 
control action considered in the delay-free case, where the additional
superscript indicates the presence of delays.

The triggering instants are now given by
$t^i_{k+1} = s^i_{k} + \Delta^i_k$, where 
\begin{eqnarray} \label{eq:sam_delay}
\Delta^i_k:=
\def\arraystretch{2.2}
\left\{
\begin{array}{ll}
\displaystyle \frac{|\ave^{w,\tau}_i(s^i_{k})|}{4d_{i}} & \quad \textrm{if } \, 
|\ave^{w,\tau}_i(s^i_{k}) |\ge \eps_i(s^i_{k})  \\
\quad\ \displaystyle \frac{\eps}{4d_{i}} & \quad
\textrm{otherwise} 
\end{array} \right.
\end{eqnarray}
which is also the natural generalization of the 
triggering rule considered in the delay-free case.
As before, by construction the inter-sampling times 
are bounded away from zero. Notice that by construction
$s^i_k \geq t^i_k$ with equality holding if and only if delays are zero,
and $t^i_{k+1} > s^i_k$.

Approaching the analysis directly with respect to $\ave^{w,\tau}_i$ is not 
simple because $\ave^{w,\tau}_i$ contains data which are collected at  
different time instants. Nonetheless, one can simplify the analysis
by exploiting the special structure of the control law. Rewrite
{\setlength\arraycolsep{2pt}
\begin{eqnarray} \label{eq:state_noisy_delay_equiv}
z_j(v_k^{ij}) &=& x_j(v_k^{ij}) + w_j(v_k^{ij}) \nonumber \\
&=& x_j(s_k^{i}) + \bar w_{ij}(s_k^{i})
\end{eqnarray}}%
where 
{\setlength\arraycolsep{2pt}
\begin{eqnarray} \label{eq:noise_delay_equiv}
\bar w_{ij}(s_k^{i}) := w_j(v_k^{ij}) + x_j(v_k^{ij}) - x_j(s_k^{i})
\end{eqnarray}}%
Since the control action does always belong to $\{-1,0,1\}$ and since 
$s_k^{i} -v_k^{ij} \leq s_k^{i} -t_k^{i} \leq \max_{j\in\neigh{i}} \tau_k^{ij}$, 
we are guaranteed that
$|\bar w_{ij}(s_k^{i})| \leq |w|_\infty + \tau_{max}$,
where
\begin{eqnarray} \label{eq:total_delay}
\tau_{max} := \sup_{k \in \mathbb N_0}  \max_{i \in I} \max_{j \in \mathcal N_i} \tau_k^{ij}
\end{eqnarray}
represents the maximum delay that can occur over a network communication link. 
It follows that
{\setlength\arraycolsep{2pt}
\begin{eqnarray} \label{eq:ave_noisy_delay_equiv}
\ave^{w,\tau}_i(s^i_k) &=& \sum_{j\in\neigh{i}}(x_j(s_k^{i})-x_i(s^i_k)) + \sum_{j\in\neigh{i}} \bar w_{ij}(s_k^{i}) \nonumber \\
&=& \ave_i(s^i_k) + \sum_{j\in\neigh{i}} \bar w_{ij}(s_k^{i})
\end{eqnarray}}%
This suggests that the analysis for the case of delays can be approached as in the delay-free case
by considering a different, possibly larger, noise contribution.

The first result is concerned with boundedness of the state trajectories,
and is a straightforward variation of Theorem \ref{thm:statebounded}.

Let
\begin{eqnarray}
\bar \gamma:=\left(\frac{1}{3}+\frac{4}{3} \frac{d_{max}}{\eps}\right) ( |w|_\infty + \tau_{max})
 \label{eq:gamma_delay}
\end{eqnarray}
 
 \begin{theorem}\label{thm:statebounded_delay}
	Consider a network of $n$ dynamical systems as in (\ref{eq:modelloA-cont}), which are
	interconnected over an undirected connected graph $G = (I,E)$. 
	Let each local control input be generated in 
	accordance with (\ref{eq:controls_delay})-(\ref{eq:sam_delay}). Then, for every initial condition, the state $x$ satisfies
	\begin{eqnarray}\label{eq:xsup_delay}
	\max_{i\in I} x_i(t) \leq 
	\left\{
	\def\arraystretch{1.2}
	\begin{array}{rl}
	\overline x & \quad \mathrm{if}\ |\overline x|\ge \bar \gamma\\ 
	\bar \gamma & \quad \mathrm{otherwise}
	\end{array} \right.
	\end{eqnarray}
	and
	\begin{eqnarray}\label{eq:xinf_delay}
	\min_{i\in I} x_i(t) \geq 
	\left\{
	\def\arraystretch{1.2}
	\begin{array}{rl}
	\underline x & \quad \mathrm{if}\ |\underline x|\ge \bar \gamma\\ 
	- \bar \gamma & \quad \mathrm{otherwise}
	\end{array} \right.
	\end{eqnarray}
	for every $t \in \mathbb R_{\geq 0}$.
\end{theorem} 

\emph{Proof.}  The proof follows exactly the same steps as the 
proof of Theorem \ref{thm:statebounded} using condition 
$x_i(s^i_k) > \overline x - \frac{1}{3} |w|_\infty - \frac{1}{3} \tau_{max}$
for Sub-case 1 and condition $x_i(s^i_k) \leq \overline x - \frac{1}{3} |w|_\infty - \frac{1}{3} \tau_{max}$
for Sub-case 2. \qedp
 \smallskip 

The counterpart of Theorem \ref{thm:generalnoise} is slightly more involved but
it essentially follows the same reasoning of Section V-C.

Let
\begin{eqnarray} \label{eq:r_delay}
\bar r:=\max\{\eps,\eps\chi_0\}+\left(\frac{\eps}{3}+3d_{max}\right)
( |w|_\infty + 3 \tau_{max} )
\end{eqnarray}

\begin{theorem} \label{thm:generalnoise_delay}
Consider a network of $n$ dynamical systems as in (\ref{eq:modelloA-cont}), which are
interconnected over an undirected connected graph $G = (I,E)$. 
Let each local control input be generated in accordance with (\ref{eq:controls_delay})-(\ref{eq:sam_delay}). 
Assume that noise and delays are such that $\eps \leq 2 d_{max} (|w|_{\infty} + 3 \tau_{max})$.
Then, for every initial condition, the network state $x$ 
enters in a finite time the set 
\begin{eqnarray}\label{eq:setD_delay}
{\bar {\cal D}}:=\{x\in\mathbb{R}^n: |\sum_{j\in {\cal {N}}_i}(x_j-x_i)| < \bar r,\ \forall i\in I\}
\end{eqnarray} 
and remains there forever. 
\end{theorem} 
 \smallskip
  
The proof of Theorem \ref{thm:generalnoise_delay} hinges upon two 
technical results, which extend Lemma \ref{lem:elessL} and \ref{lem:samesign}
to the presence of delays.

\begin{lemma}\label{lem:elessL_delay}
Consider the same assumptions and conditions as in Theorem \ref{thm:generalnoise_delay}. 
For any $i\in I$, it holds that
\begin{eqnarray}
\eps_i(s_k^i) \leq \bar r-\frac{5}{3}d_{max} ( |w|_{\infty} + 3 \tau_{max} )
\end{eqnarray}
for every $k \in\mathbb{N}_0$. \smallskip
\end{lemma}

\emph{Proof.} By Theorem \ref{thm:statebounded_delay}, we have 
 \begin{eqnarray}
  |x_i(s_k^i)| \leq \max \{|\overline x|,|\underline x|,\bar \gamma\}\le \chi_0+ \bar \gamma
 \end{eqnarray}
 Hence,
{\setlength\arraycolsep{2pt}
 \begin{eqnarray}
 \eps_i(s_k^i) % &=&\max\{\eps,\eps|x_{i}(s_k^i)|\}\nonumber\\
 &\le& \max\{\eps,\eps(\chi_0+\bar \gamma)\}\nonumber\\
 &\leq& \max\{\eps,\eps\chi_0\}+\eps \bar \gamma \nonumber\\
 &\leq& \bar r-\frac{5}{3}d_{max} ( |w|_\infty + 3 \tau_{\max} )
 \label{eq:lem1_delay}
 \end{eqnarray}}%
where the last inequality holds by the definitions \eqref{eq:r_delay} 
and \eqref{eq:gamma_delay} of $\bar r$ and $\bar \gamma$ respectively.\qedp

 \begin{lemma} \label{lem:samesign_delay} 
 	Consider the same assumptions and conditions as in Theorem \ref{thm:generalnoise_delay}.
 	Consider any index $i \in I$ and any 
	$M \in \mathbb N_0$.
	If $|\ave_{i} (s^i_{k+m})| \geq \bar r$ for $m = 0,1,\ldots,M$
	then
 	\begin{eqnarray}
    \sign(\ave_{i} (s^i_{k+m})) = \sign(\ave_{i} (s^i_{k})), \nonumber \\
    m = 1,2,\ldots,M+1
 	\end{eqnarray}
 \end{lemma}

 \emph{Proof.}  Assume without loss of generality that $\ave_{i} (s^i_{k})\ge \bar r$,
 the other case being analogous. From Lemma \ref{lem:elessL_delay}, we have
 {\setlength\arraycolsep{2pt}
 \begin{eqnarray}
 \ave_{i}^{w,\tau}(s_k^i)& \geq &\ave_{i} (s^i_{k})-d_{max} ( |w|_\infty + \tau_{\max} ) \nonumber\\
 & \geq & \bar r - d_{max} ( |w|_\infty + \tau_{\max} ) \nonumber\\
 & \geq & \eps_i(s_k^i)
\end{eqnarray}}%
Hence, $u_{i} (s^i_{k}) = 1$. Moreover,
{\setlength\arraycolsep{2pt}
\begin{eqnarray} \label{}
\ave_{i} (t) &\geq& \ave_{i} (s^i_k)  -2 d_i (s^i_{k+1} -s^i_k)  \nonumber \\
&=& \ave_{i} (s^i_k)  -2 d_i (t^i_{k+1} -s^i_k) -2 d_i (s^i_{k+1} -t^i_{k+1})  \nonumber \\
&\geq& \ave_{i} (s^i_k) - \frac{1}{2} \ave^{w,\tau}_{i} (s_k^i) - 2 d_{max} \tau_{\max} \nonumber \\
% &=& \frac{1}{2}  \ave_{i} (s^i_k) - \frac{1}{2} \sum_{j\in\neigh{i}} \bar w_{ij}(s_k^{i}) -  2 d_{max} \tau_{\max} \nonumber \\
&\geq& \frac{1}{2}  \ave_{i} (s^i_k) - \frac{1}{2} d_{max} |w|_\infty - \frac{5}{2}  d_{max} \tau_{\max} \nonumber \\
&\geq& \frac{1}{2} \bar r - \frac{1}{2} d_{max} |w|_\infty - \frac{5}{2}  d_{max} \tau_{\max} \nonumber \\
&>&  \frac{1}{2} \max\{\eps,\eps\chi_0\} 
\end{eqnarray}}% 
for all $t \in [s^i_k,s^i_{k+1}]$. The first inequality comes from the fact that 
the control inputs always belong to $\{-1,0,1\}$. Thus, over the time interval 
$[s^i_k,s^i_{k+1}]$, the value of $\ave_{i}$ can decrease at most with slope $2 d_i$.
The second inequality follows because $s^i_{k+1} -t^i_{k+1} \leq \tau_{max}$.

We then conclude that $\ave_{i} (s^i_{k+1}) >0$. 
Thus $\ave_{i}$ preserves its sign. \qedp
 \smallskip
 
We can now proceed with the proof of Theorem \ref{thm:generalnoise_delay}. 
 \smallskip
    
\emph{Proof of Theorem \ref{thm:generalnoise_delay}.} 
Like in the proof of Theorem \ref{thm:generalnoise}, we 
introduce three sets into which we partition the set of switching times of each node $i$.
For each $i \in I$, let   
{\setlength\arraycolsep{1pt} 
\begin{eqnarray}
\def\arraystretch{1.5}
\begin{array}{l}
\mathscr W_{i1} := \big\{ s^i_k  : \,  |\ave^{w,\tau}_{i} (s^i_k)| \geq \eps_{i} (s^i_k)   \wedge  
 |\ave_{i} (s^i_k)| \geq \bar r \big\}   \\ 
\mathscr W_{i2} := \big\{ s^i_k  :  \, |\ave^{w,\tau}_{i} (s^i_k)| \geq \eps_{i} (s^i_k) \wedge
|\ave_{i} (s^i_k)| < \bar r  \big\}  \\
\mathscr W_{i3} := \left\{ s^i_k  : \,  |\ave^{w,\tau}_{i} (s^i_k)|
< \varepsilon_{i} (s^i_k) \right\} 
\end{array}  
\end{eqnarray}}%
Clearly, $t^i_k \in \mathscr S_{i1} \cup \mathscr S_{i2} \cup \mathscr S_{i3}$ 
for every $k \in \mathbb N_0$.

Pick any $i \in I$, and assume by contradiction that there exists a time $t_*$ such that
$|\ave_{i} (s^i_{k})| \geq \bar r$ for all $s^i_{k}\geq t_*$. 
In view of Lemma \ref{lem:elessL_delay}, $u_i$ is
never zero from $t_*$ on since the condition above yields 
$|\ave^{w,\tau}_{i} (s^i_{k})| \geq \bar r - d_{max} ( |w|_\infty + \tau_{max} ) \geq \eps_i(s^i_k)$.
Moreover, by Lemma \ref{lem:samesign_delay}, $\sign(\ave_{i} (s^i_{k+m})) = \sign(\ave_{i} (s^i_k))$ for every $m$. 
Hence, either $u_i(t) = 1$ for all $s^i_{k} \geq t_*$ 
or $u_i = -1$ for all $s^i_{k} \geq t_*$. This would imply that $x_i$ diverges,
violating the state boundedness property of Theorem \ref{thm:statebounded_delay}.

 By the foregoing arguments, there exists a time instant $s^i_k$ such that 
 $|\ave_{i} (s^i_{k})| < \bar r$. This implies that $s^i_{k} \notin \mathscr W_{i1}$, {or, equivalently, 
 that $s^i_{k} \in \mathscr W_{i2}\cup \mathscr W_{i3}$.} 
 Thus it remains to show that transitions from $\mathscr W_{i2}$ and $\mathscr W_{i3}$ 
 to $\mathscr W_{i1}$ are not possible. We analyze the two cases separately.

\emph{Case 1: $s^i_{k} \in \mathscr W_{i2}$.}
In this case, $u_i(s^i_k) = \{-1,1\}$. 
Suppose that $u_i(s^i_k)=1$, 
the other case being analogous. Then, 
{\setlength\arraycolsep{2pt} 
\begin{eqnarray}
\ave_{i} (t) \leq \ave_{i} (s^i_k) < \bar r 
\end{eqnarray}}%
for all $t \in [s^i_k,s^i_{k+1}]$,
where the first inequality follows since $u_i(s^i_k)=1$
and the second because $s^i_k \in \mathscr W_{i2}$ by hypothesis. 
Moreover, $u_i(s^i_k)=1$ implies $\ave^{w,\tau}_{i} (s^i_k) \geq \varepsilon_{i} (s^i_k)$
so that $\ave_{i} (s^i_k) \geq \varepsilon_{i} (s^i_k) - d_{max} (|w|_\infty + \tau_{\max})$.
Hence, 
{\setlength\arraycolsep{2pt} 
\begin{eqnarray} \label{eq:pointwise_invariance_1_delay}
\ave_{i} (t) &\geq& \ave_{i} (s^i_k) - 2d_i (t^i_{k+1}-s^i_k) - 2 d_i (s^i_{k+1}-t^i_{k+1})  \nonumber \\
&=& \ave_{i} (t^i_k) -  \frac{1}{2} \ave^{w,\tau}_{i} (s^i_k)  - 2 d_{max} \tau_{\max}  \nonumber \\
% &=& \frac{1}{2}  \ave_{i} (s^i_k) - \frac{1}{2} \sum_{j\in\neigh{i}} \bar w_{ij}(s_k^{i}) -  2 d_{max} \tau_{\max} \nonumber \\
&\geq& \frac{1}{2}   \ave_{i} (s^i_k) -  \frac{1}{2} d_{max} |w|_\infty -  \frac{5}{2} d_{max} \tau_{\max}  \nonumber \\
&\geq& \frac{1}{2}  \varepsilon_{i} (t^i_k) - d_{max} |w|_\infty  - %\frac{7}{2}  
{3} d_{max} \tau_{\max} \nonumber \\
&> &  - d_{max} |w|_\infty  -  %\frac{7}{2} 
{3}
d_{max} \tau_{\max}  \nonumber \\
&>& - \bar r  
\end{eqnarray}}%
for all $t \in [s^i_k,s^i_{k+1}]$. 
Hence, $|\ave_{i} (s^i_{k+1})| < \bar r$ which implies that $s^i_{k+1} \notin \mathscr W_{i1}$.
 
\emph{Case 2: $s^i_{k} \in \mathscr W_{i3}$.}
In this case we have ${u_i}(t)=0$ for all $t\in[s_k^i, s_{k+1}^i]$. 
Since $u_i(s^i_k)=0$ then $|\ave^{w,\tau}_{i} (s^i_k)| < \varepsilon_{i} (s^i_k)$
so that $|\ave_{i} (s^i_k)| < \varepsilon_{i} (s^i_k) + d_{max} (|w|_\infty + \tau_{\max})$.
Moreover, $t^i_{k+1}-s^i_k= \varepsilon/(4d_i)$.
Hence,
{\setlength\arraycolsep{2pt} 
 \begin{eqnarray}\label{eq:avethm4p_delay}
 |\ave_{i} (t)| &\leq& |\ave_{i} (s^i_k)| + d_i (t^i_{k+1}-s^i_k) + d_i (s^i_{k+1}-t^i_{k+1})  \nonumber \\
 &<& \eps_{i} (t^i_k) + d_{max} ( |w|_\infty + \tau_{max} ) +\frac{\eps}{4} + d_{max} \tau_{max} \nonumber \\
 &\leq& \eps_{i} (t^i_k) +\frac{3}{2} d_{max}|w|_\infty + 3 d_{max} \tau_{max}   \nonumber \\
 &<& \bar r
 \end{eqnarray}}%
{for all $t \in [s^i_k,s^i_{k+1}]$,}
where the {third} inequality 
follows from $\eps\le 2d_{max} ( |w|_\infty + 2 \tau_{max})$ and 
the fourth one follows from Lemma \ref{lem:elessL_delay}. Hence, $s^i_{k+1} \notin \mathscr W_{i1}$.

Hence, we conclude that 
$s^i_\ell\in \mathscr W_{i2} \cup \mathscr W_{i3}$ for all $\ell\ge k$. Moreover, the previous arguments 
show that $ |\ave_{i} (t)|<r$ for all $t\in [s^i_\ell,s^i_{\ell+1}]$, for all $\ell\ge k$, 
which guarantees that $x$
remains forever inside $\bar {\mathcal D}$. 
 \qedp 
 \smallskip
 
Following the same steps as in Section V-B, it is an easy matter to 
see that if $\eps > 2 d_{max} (|w|_{\infty} + 3 \tau_{max})$ then the state $x$
converges in a finite time to a point belonging to the set $\bar {\mathcal{D}}$,
which parallels the result in Theorem \ref{thm:smallnoise}.

We close this section by pointing out that less conservative bounds can
be obtained under the additional hypothesis that the messages are 
time-stamp synchronized, in which case one can assume that if node $i$ sends a request to node $j$ 
at time $t^i_k$ then $j$ is capable of providing node $i$ with the value $z_j(t^i_k)$.
This scenario has been studied in \cite{de2013robust}.

\bibliographystyle{IEEEtran}
\bibliography{paper_biblio}

%%%%%%%%%%%%%%%%%%%%%%%%%%%%%%%%%%%
%%%%%%%%%%%%%%%%%%%%%%%%%%%%%%%%%%%%%%%
\end{document}